\documentclass[fleqn,usenatbib]{mnras}

\usepackage{newtxtext,newtxmath}

\usepackage[T1]{fontenc}

\DeclareRobustCommand{\VAN}[3]{#2}
\let\VANthebibliography\thebibliography
\def\thebibliography{\DeclareRobustCommand{\VAN}[3]{##3}\VANthebibliography}
\newcommand\HII{H{\sc ii}}
\newcommand\HI{H{\sc i}}

\usepackage{graphicx}	
\usepackage{amsmath}	
\usepackage{amssymb}	

\title[Tol 1004-296 and Tol 0957-278 Kinematics Connection]{The influence of ionized gas kinematics on \HII\ galaxies \\
The cases of Tol 1004-296 and Tol 0957-278}

\author[H. Plana]{
Henri Plana,$^{1}$\thanks{E-mail: plana@uesc.br}
Vitor G. Alves,$^{1}$
Maiara S. Carvalho$^{2}$
\\
$^{1}$Laborat\'orio. de Astrof\'{i}sica Te\'orica e Observacional - Departamento de Ciências Exatas - Universidade Estadual de Santa Cruz, 45662-900 Ilh\'eus-BA, Brazil\\
$^{2}$Instituto de Astronomia, Geof\'{i}sica e Ci\^encias Atmosf\'ericas  Cidade Universit\'aria, Rua do Mat\~ao, 1226 - S\~ao Paulo-SP, Brazil \\
}

\date{Accepted XXX. Received YYY; in original form ZZZ}

\pubyear{2015}

\begin{document}
\label{firstpage}
\pagerange{\pageref{firstpage}--\pageref{lastpage}}
\maketitle

\begin{abstract}
Blue Compact Galaxies (BCGs), also known as \HII\ galaxies, are dwarf, star-forming objects with relatively simple dynamics, which allows for the investigation of star formation mechanisms in a cleaner manner compared to late-type objects. In this study, we have examined various characteristics of the interstellar medium, in connection with the 
kinematics and dynamics of ionized gas, in Tol 1004-296 and Tol 0957-278.
These two objects were observed using the SOAR Integral Field Spectrometer (SIFS) attached to the Southern Observatory for Astrophysical Research (SOAR). Both galaxies were observed with two gratings: one with medium resolution for monochromatic and abundance maps, and another with high resolution for kinematics and profile analysis. Additionally, we conducted an analysis on the velocity and velocity dispersion maps using intensity-velocity dispersion (I - $\sigma$) and velocity-velocity dispersion (Vr - $\sigma$) diagrams.
Neither object exhibits a rotation pattern, and only Tol 1004-296 shows a velocity gradient between the two principal knots. However, the study reveals the significant role played by  velocity dispersion in the star formation process. Specifically, we identified a relationship between monochromatic intensity, metallicity, and velocity dispersion, where high emission corresponds to low metallicity and low velocity dispersion. Tol 1004-296, in particular, exhibits a distinctive linear high velocity dispersion pattern between the two main knots, suggesting that both star formation sites are pushing the gas in opposite directions.
\end{abstract}



\section{Introduction}          

The concept of nuclear activity and starburst galaxies, characterized by their blue color, has been known since pioneering work by \citet{Haro1956}, \citet{Ambartsumian1968}, \citet{Zwicky1964}, \citet{Zwicky1966}, and \citet{Markarian1967}. These objects collectively represent a diverse array of galaxies, each with unique physical and morphological characteristics. However, soon after these initial studies, a classification scheme began to emerge in the literature.
Criteria such as luminosity and morphological properties were utilized to classify the galaxies observed in the aforementioned surveys. The term "Blue Compact Galaxies" (BCG), as defined by \citet{Zwicky1971}, refers to objects (or parts of objects) with surface brightness brighter than 20 mag arcsec$^{-2}$ in both blue and red plates. \citet{Thuan1981} later referred to these objects as Blue Compact Dwarfs (BCDs), defining them as having an absolute blue magnitude fainter than M$_B = -18.15$ mag, a diameter less than 1 kpc, and exhibiting strong emission lines superimposed on a blue continuum.
Furthermore, the spectra of these objects can also serve as selection criteria. \citet{Sargent1970} published findings on "isolated extragalactic \HII\ regions" with spectra resembling \HII\ regions, characterized by prominent strong emission lines above a weak stellar continuum, and morphologies consistent with BCDs. \citet{Terlevich1981} further investigated extragalactic \HII\ regions, examining their chemical composition, luminosity, and velocity dispersion.

The entities now termed \HII\ galaxies began to be studied more systematically in the 1990s. \citet{Terlevich1991} and \citet{Salzer1989a, Salzer1989b} presented imaging and spectrophotometric catalogs of these galaxies. The morphology of \HII\ galaxies is diverse; while many exhibit an irregular shape, some display spiral arms and multiple bright knots indicative of intense star formation sites \citep{Telles1997a, Mendez2000, Lagos2011}.

In addition to their low luminosity ($M_B > -18$ mag) and compactness (diameter < 1 kpc), \HII\ galaxies also demonstrate low metallicity (between 1/50 and 1/3 of solar metallicity) \citep{Thuan1985}.

Initially, due to their low metallicity and high EW(H$\alpha$), BCDs were believed to be young galaxies in the process of forming their first generation of stars. However, this hypothesis was swiftly challenged. Broadband optical and near-infrared imaging and photometry revealed the presence of an underlying older stellar population, suggesting that the intense star-forming episodes (starbursts) were merely phases followed by more prolonged and quiescent periods \citep{Thuan1983, Telles1995, Papaderos1996a, Papaderos1996b, Telles1997a, Cairos2003, MunozMateos2009}. \citet{Westera2004} and \citet{Cuisinier2006} further investigated these distinct stellar populations: young, intermediate, and old.

Even though it is becoming clear that BCDs or \HII\ galaxies are not early galaxies forming their first stars, they still offer certain advantages that make them the best objects for studying Star Formation (SF) mechanisms \citep{Cairos2015}. Their relatively simple dynamics, without spiral arms resulting from density waves and absence of disk instabilities, allows for the investigation of SF mechanisms in a cleaner manner compared to late-type objects \citep{Hunter2004}. The shallow potential well also facilitates feedback processes \citep{MacLow1999}. High spatial resolution (approximately $2-12$ pc) hydrodynamical simulations indicate that starbursts result from gas fragmentation into massive and dense clouds rather than large-scale inflows of gas \citep{Perret2014}, although this remains a topic of ongoing debate. Often referred to as "local Building Blocks," BCGs offer our best chance to understand the properties of high-redshift galaxies \citep{Kunth2000}, and they serve as ideal laboratories for studying star formation processes in massive clusters ($>10^4 M_\odot$) \citep{Greis2016, Brennan2017}.

The emergence of Integral Field Units (IFUs) in the 2000s has allowed for the mapping of electronic density, reddening, metallicity, ionization mechanisms, and gas kinematics of these objects. Depending on the instrument used, a trade-off between field of view and spatial sampling must be made. Over the last 15 years, several studies have been conducted using IFUs. Among them, \citet{GarciaLorenzo2008}, \citet{Lagos2009}, \citet{Bordalo2009}, \citet{Castillo-Morales2011}, \citet{Cairos2015}, \citet{Cairos2017a}, \citet{Cairos2017b}, \citet{Carvalho2018}, \citet{Cairos2020}, \citet{Egorov2021} and \citet{Cairos2022} have investigated more than 15 objects using IFU instruments in various ways. The most recent studies have focused on peculiar objects such as Mrk 900, Haro 14, and DDO 53.

These studies reveal some common characteristics. Most of these galaxies are unrelaxed objects, displaying irregular morphologies with clumpy SF sites and filamentary features of the ionized gas. Nearly all objects exhibit strong differences between the gas and stellar distributions. Several galaxies contain Wolf-Rayet (WR) stars \citep{Conti1990}. The ionization mechanisms predominantly stem from photoionization, although some objects (less than 50\%) also show shock contributions in the outer regions. Extinction maps exhibit inhomogeneity. In some cases, the velocity field displays a rotation pattern (or at least a coherent velocity gradient), and the velocity dispersion maps are also inhomogeneous. Some studies of individual objects reveal two stellar populations as mentioned earlier: a very young 5.5-6.5 Myr stellar population in the SF sites and a Gyr-old population in an extended Low Surface Brightness (LSB) part of the galaxy.

The formation of BCD galaxies remains a subject of debate, with one scenario proposing interaction/fusion between low-mass, gas-rich galaxies. \citet{Chhatkuli2023} studied a sample of BCDs and found traces of dwarf-dwarf mergers. Particularly, they demonstrated the presence of a young stellar population (a few 10 Myrs old) as a result of the merger. A detailed study of a specific object, VCC 848, utilizing N-body/hydrodynamical simulations, \HI, H$\alpha$, and Herschel PACS observations, revealed that this object is the outcome of a merger between a gas-dominated progenitor and a gas-bearing star-dominated galaxy \citep{Zhang2020}.

In this paper, we investigate two \HII\ galaxies, Tol 1004-294 and Tol 0958-278, using the SIFS-IFU at the SOAR Telescope. We present a more detailed gas kinematics analysis of these objects to shed new light on the star formation mechanisms.

The paper is organized as follows: Section 2 presents the data, a description of the objects, and details of the data reduction. Section 3 describes the different results obtained. Finally, Section 4 concludes the article.

\begin{figure*}
	\includegraphics[scale=0.35]{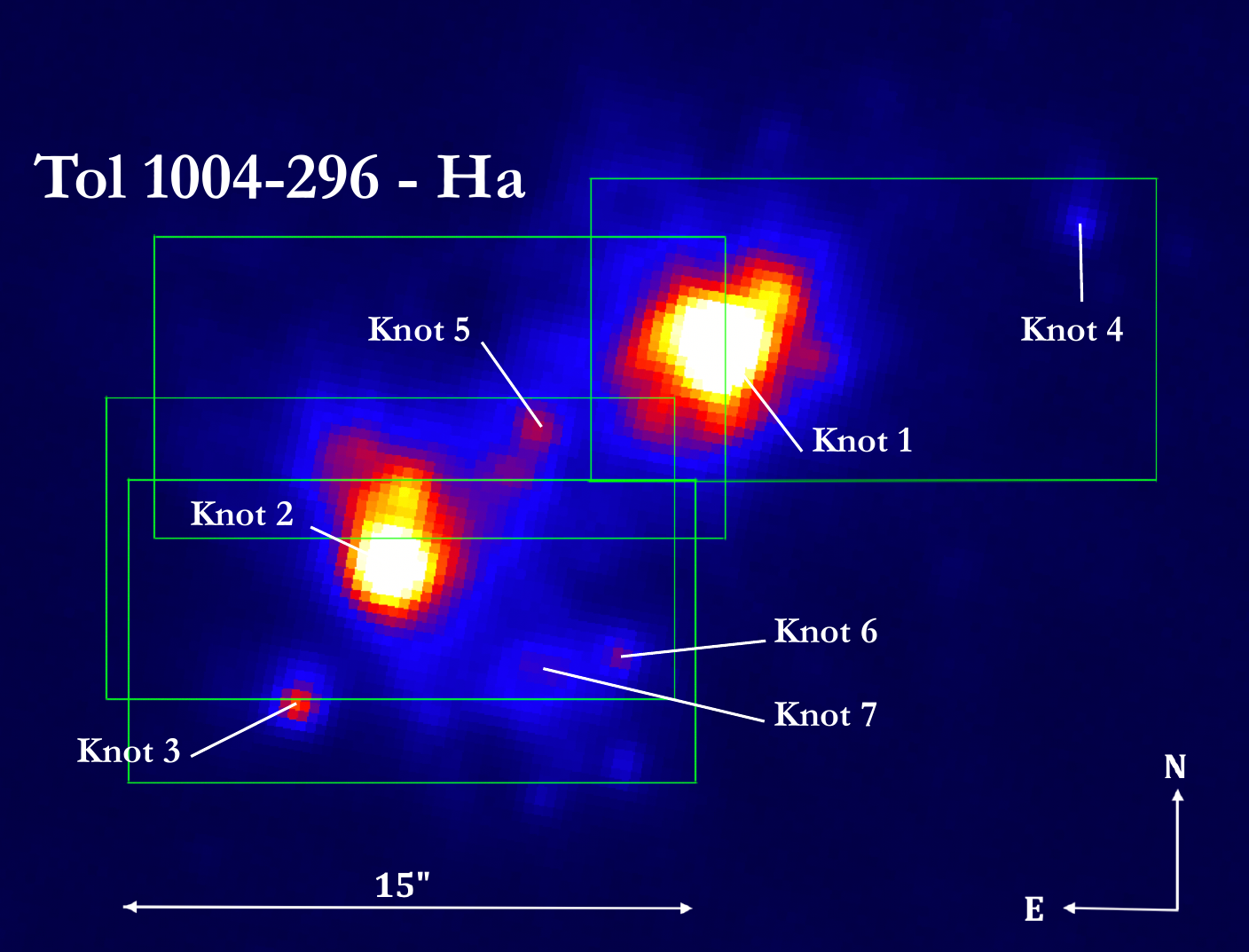}\quad\includegraphics[scale=0.35]{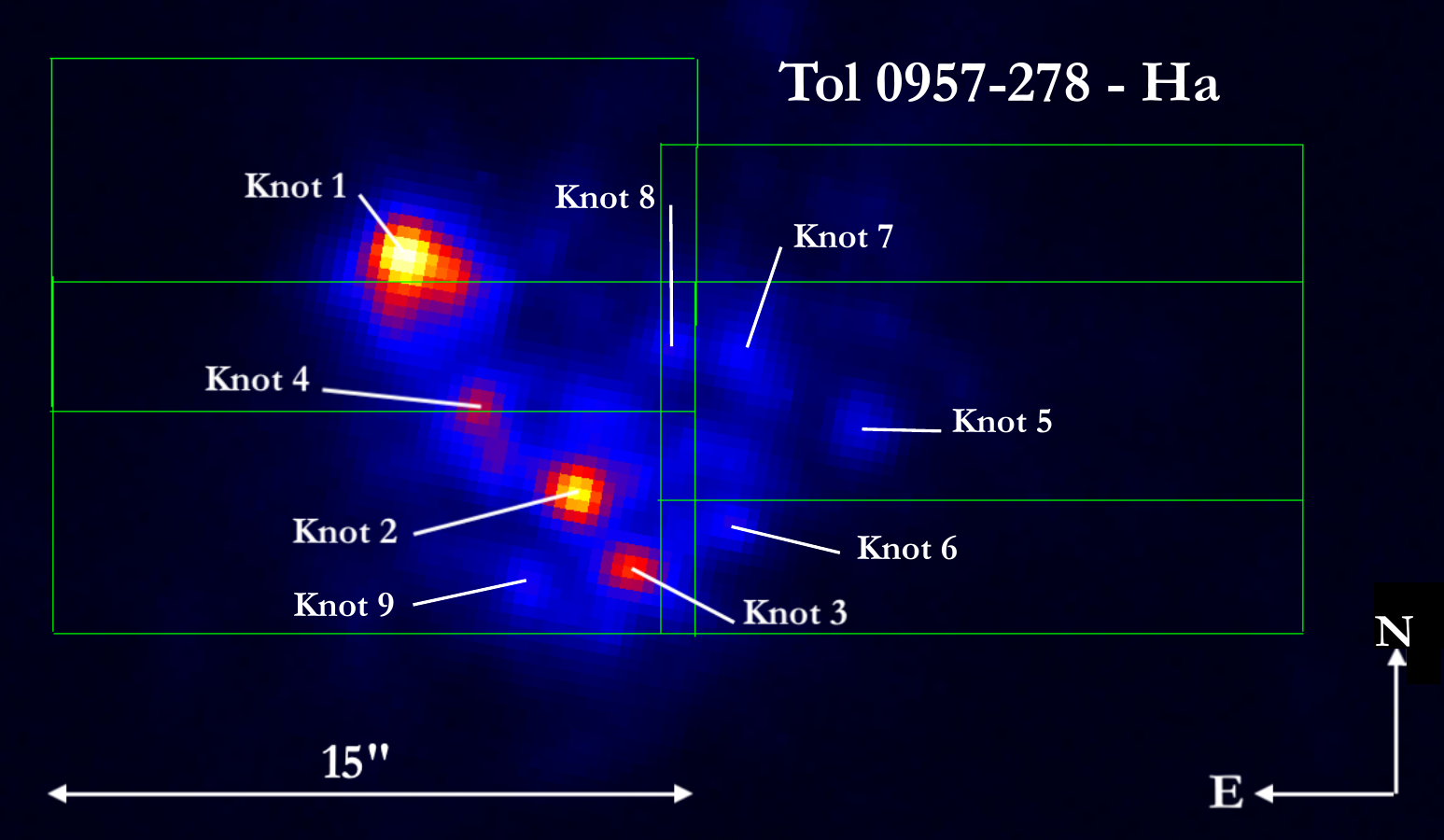}
    \caption{H$\alpha$ images of Tol 1004-296 and Tol 0957-278 from the Irénée du Pont telescope at LCO Observatory.} 
    \label{fig:LTO_Tol1004_Tol0957}
\end{figure*} 

\section{The data}        

We are presenting here two galaxies, Tol 1004-296 and Tol 0957-278 (also known as NGC 3125 and Tol 002). They are catalogued as Blue Compact Dwarf objects \citep{GildePaz2005}, also commonly called H{\sc ii} galaxies \citep{Terlevich1991}. These objects have been chosen following the criteria: (1) Presence of strong emission lines with a flux limit of 5.0 $\times$ 10$^{-15}$ erg s$^{-1}$ cm$^{-2}$ (for [NII]6584\AA) in order to have an SNR no lower than 3; (2) Low redshift in order to keep good spatial sampling; (3) Large enough in order to have the maximum number of pixels for the diagnostic diagrams (but not too large in order to be observed with a reasonable number of IFU fields); (4) Integrated velocity dispersion larger than 30 km s$^{-1}$ \citep{Bordalo2011}.

Both galaxies are characterised by a strong gas concentration and a high star formation rate, with a spectrum dominated by intense emission lines superimposed on a weak stellar continuum \citep{Terlevich1991}.

\subsection{Tol 1004-296}

The dwarf galaxy Tol 1004-296, also known as NGC 3125, ESO 435-G 041, ESO 100419-2941.5, AM 1004-294, MCG 05-24-022, is classified as a BCG with a magnitude of $M_V$ = -17.7 \citep{Bergvall1986} and a low metallicity of 12 + Log(O/H) = 8.34 \citep{Vacca1992}. It is located at a distance of 11.99 Mpc \footnote{Local Group distance from Nasa Extragalactic Database}. \citet{Smith1976} described it as an object with super condensations showing emission lines and a diffuse stellar continuum, from an objective prism survey \citep{Smith1975}. Spectroscopic survey by \citet{Penston1977} confirmed these objective prism data, exhibiting HeII$\lambda$4686 emission line in one of the condensation.
\citet{Reif1982} has detected \HI\ content and \citet{Paturel2003} converted it to mass according to \citet{Pustilnik2007}, giving an $M_{HI} = 3.19 \times 10^8 M_{\odot} $.
\citet{Kunth1985} report the presence of {\it Wolf Rayet} (WR) stars in the object.
Dust content and mass have been estimated to $4.4 \times 10^{5} M_{\odot} $ by \citet{Engelbracht2008}, using {\it Spitzer Observatory}, also giving a dust temperature of $53.7 \pm 0.3$ K.

Tol 1004-296 has been observed in the R, B and H$\alpha$ bands on several occasions. \citet{Kunth1988}, \citet{Papaderos2002}, \citet{GildePaz2003, GildePaz2005} present studies of this object based on images in these three bands. \citet{Kunth1988} B image mentioned four intense star formation regions.
\citet{GildePaz2003} shows both the broad and narrow bands images of the object.
It is notable that the extension of the H$\alpha$ emission goes beyond 40\arcsec. We can also see that this extension, at lower limits, appears as filaments. A 20\arcsec $\times$ 20\arcsec zoomed field of view is also presented in \citet{GildePaz2003} H$\alpha$ images (slightly different compared to the observed field with SIFS - see below).

The spectrum shown by \citet{Terlevich1991} is dominated by intense emission lines of hydrogen (H$\alpha$ and H$\beta$) and forbidden oxygen lines such as [OIII]$\lambda$5007 and [OIII]$\lambda$4959. \citet{Kehrig2004} also give the fluxes of the main emission lines from the high efficient spectrograph $FEROS$ using one fibre.

\citet{Lagos2007} published an H$\beta$ emission map and equivalent width (EW) image showing EW(H$\beta$) above 90\AA. Three intense emission sites are also visible in these maps.

In \citet{Schwartz2006}, low-amplitude outflows were shown through spectroscopy conducted with the HST in the two main star-forming regions of the object.

Observations in the near-infrared, using the ESO-VLT SINFONI \citep{Eisenhauer2003}, made it possible to obtain images of the continuum emission in the K-band, where a series of faint compact sources are detected near the bright central region \citep{Vanzi2011}. The SINFONI field was 8 \arcmin $\times$ 8 \arcmin with a resolution of 0.250 $\times$ 0.125 arcsec/pixel. \citet{Vanzi2011} focused specifically on node \textit{1} in the nomenclature of \citet{Lagos2007} and \textit{A} in the nomenclature of \citet{Vacca1992} and \citet{Stevens2002}.

The galaxy is also composed of starburst regions, as well as being populated by massive star clusters that span a range of ages. Additionally, \citet{Vanzi2011} found that in several areas there is thermal bremsstrahlung emission and molecular hydrogen clouds separate from the \HII\ regions. These regions of molecular hydrogen are mainly excited in a fluorescent manner, extending beyond the giant \HII\ regions and outlining the shape of the molecular cloud involved in star formation \citet{Vanzi2011}.

Another component of this galaxy is the presence of peaks of [FeII] emission located outside the \HII\ regions, indicating the presence of supernovae that may have triggered starburst episodes. \citet{Vanzi2011} suggests that these episodes occurred in a chain, with the older one triggering the most recent one. This hypothesis was also raised by \citet{Cresci2010}.

Both \citet{Cresci2010} and \citet{Vanzi2011} show regions where starburst episodes are occurring. However, \citet{Vanzi2011} found that the massive starburst episodes occurred in three separate episodes, resulting in the distinct intense emission sources.

The massive starburst episodes populate the galaxy with hot and young stars. However, this doesn't imply the presence of older stars in the galaxy: one characteristic of \HII\ galaxies is the spatial dominance of young stars, making it challenging to detect older stellar populations \citep{Westera2004}.

Tol 1004-296 exhibits the strongest known emission in [HeII]$\lambda 1640$ of stellar origin, providing important information about the most massive stars in the galaxy. In \citet{Chandar2004}, a study of the massive stellar content (Wolf-Rayet stars and O stars) in Tol 1004-296 was conducted, revealing a significant predominance of WN late-type Wolf-Rayet stars.
\citet{Westera2004} and \citet{Cuisinier2006} have estimated characteristics of young, intermediate, and old stellar populations. The ratio between young + intermediate to old stellar population mass is 1:100 and the age is between 10 to 20 Myrs. And the young stellar mass to the total stellar mass ratio is 0.008. The young population age is between 1 to 2 Myrs and the old population to the total stellar mass ratio is 0.99.

\subsection{Tol 0957-278}

The galaxy Tol 0957-278, also known as Tol 02, ESO 435-IG 020, AM 0957-275, PGC 028863, was initially observed by \citet{Smith1976} and classified as a medium-strong emission line emitter. \citet{Kunth1985, Schaerer1999, Mendez2000} also report the presence of Wolf-Rayet (WR) stars, similar to Tol 1004-296.
Broadband CCD imaging from \citet{Kunth1988} and \citet{Doublier1999} confirms the clumpy shape of the object.
\citet{Campbell1988}'s modeling confirms a low metallicity of 12 + log(O/H) = 8.1, consistent with what has been reported in more recent studies \citep{Vacca1992, Masegosa1994, Buckalew2005, Engelbracht2008}.

\citet{Terlevich1991} and \citet{Masegosa1994} published emission line fluxes and ion abundances for different samples of \HII\ galaxies, including both objects in our study. A more recent study by \citet{Kehrig2004} also provides emission line fluxes, which we used for the selection of our two targets.

Tol 0957-278 is neighbor to several other \HII\ galaxies as mentioned by \citet{Telles1995} and seems to form a pair system with ESO 435-G 016.
\citet{Sung2002} saw this pair as an interacting system.
\citet{Kim2017} studied the \HI\ content of this system, estimating Tol 0957-278's extended \HI\ gas mass to be $1.11 \times 10^8 M_{\odot}$. They also concluded that the system has been tidally disturbed, and the interaction is responsible for the recent star formation activities.

Tol 0957-278 has a distance of 9.93 Mpc\footnote{Local Group distance from NASA Extragalactic Database} and an angular extension of $\approx~ 60\arcsec \times 40\arcsec$ \citep{Torres2017}. It is also classified as a BCD galaxy due to its low B-band magnitude $M_{B}$ = -16.26 mag \citep{Kunth1988, Doublier1999, GildePaz2003}.
\citet{GildePaz2003} observed it in the H$\alpha$ band, and, similar to the previous object, Tol 0957-278 shows extended H$\alpha$ emission (25\arcsec) with a filamentary aspect on the outskirts.

\citet{Lagos2007} also published H$\beta$ emission and equivalent width (EW) images showing EW(H$\beta$) above 150\AA. Four intense emission sites are also visible in these maps.

An underlying older stellar population mixed with a much younger stellar population has been known for quite some time, but it is still difficult to separate \citep{Westera2004}. In their study, \citet{Westera2004} and \citet{Cuisinier2006} found the following characteristics of this older population. The mass ratio between the young + intermediate population/old population is 1 to 100, the mass ratio between the young population and the total stellar mass is between 0.0081 and 0.0161. Finally, they estimated that the ratio between the old population and the total stellar mass is 0.96. The age of the young population is between 1 and 2 Myrs, and the age of the intermediate population is between 50 and 200 Myrs.

\citet{Torres2017} conducted a photometric study of the object in optical, NIR broadbands filters, and narrow H$\alpha$, [OIII]$\lambda$5007 filters. They identified eight star cluster complexes (SCCs).
These SCCs are delineated by the H$\alpha$ emission, which is located in the central zone of the galaxy and extends towards the north and south regions.
They have focused on decomposing the galaxy into a disk - or host galaxy formed by old stellar populations - and the Star Formation region (Low Surface Brightness Component - LSBC and Star Formation Component). They also estimated the ages and masses of these components. The age and masses of the LSBCs span between 3.7 Myrs to 2.5 Gyrs and $2.4 \times 10^6 M_{\odot}$ to $93.3 \times10^6 M_{\odot}$. The age and masses of the SCCs span between 3.1 Myrs to 11 Myrs and $35.5 \times10^3 M_{\odot}$ to $269.1 \times 10^3 M_{\odot}$.

\begin{table}
	\centering
	\caption[Journal of Observations]{Journal of Observations \label{tab:journalobs}}
	\begin{tabular}{llcc}
\hline
\textbf{Object} & \textbf{Right}           & \textbf{Declination} & \textbf{Systemic }  \\
                       &  \textbf{Ascention}   &                                 & \textbf{Velocity$^1($km s$^{-1}$)}                                \\
\hline
Tol1004-296     & 10:06:33.30 & -29:56:05.00 & 1113  \\
Tol0957-278	& 09:59:21.50 & -28:07:58.76 & 971 \\
\hline
\hline
\textbf{Run}                   & \textbf{Grating} & \textbf{Observed} & \textbf{Exposure Time}  \\
                                     &                          & \textbf{Fields}       &  \textbf{per field}  \\
January - March 2022   & R700M            &  4                          & 2 $\times$ 20min  \\      
February 2023              &  M1500M3       &  4                          & 2 $\times$ 15min  \\
\hline
       \end{tabular}
       
\flushleft $^1$ Heliocentric Velocity from NED.
\end{table}       

\subsection{Observations}

These two objects were observed at the SOAR Observatory \footnote{Southern Observatory for Astrophysical Research} using the SIFS \citep{Lepine2003, Fraga2018} instrument \footnote{SOAR Integral Field Spectrometer}. SIFS is an Integral Field Unit (IFU) spectrograph. The instrument has a spatial scale of 0.3\arcsec~ per fiber, and the field of view is 15\arcsec $\times$ 7.8\arcsec. The fibers feed a pseudo slit to a bench spectrograph. The detector is a 4096 x 4112 pixel CCD231-84 from E2V with a pixel physical size of 15 $\mu$m$^2$. The Read-Out Noise (RON) of the detector is 2.0 e-/ADU.

Several observations were necessary to cover the targets. We used four different fields for each target resulting to a final field of view of 30\arcsec $\times$ 15 \arcsec. Each target was observed with two different gratings: the 700R (with a resolution of 4200@5500) was used for obtaining monochromatic maps for different emission lines, and the 1500M3 (with a resolution of 9500@5500) for the velocity and velocity dispersion maps. Table \ref{tab:journalobs} shows the observation log. Figure \ref{fig:LTO_Tol1004_Tol0957} illustrates the different fields using an H$\alpha$ image from the Irénée du Pont telescope at La Campanas Observatory \footnote{These images were downloaded from the NED website} for both galaxies. We also show the main emission knots.

We estimated a precision equivalent to a spaxel (0.3\arcsec) in the reconstruction of the mosaics. Observations couldn't be completed in a single semester and had to be spread out over two semesters, a year apart.

During the night, two spectrophotometric standard stars were also observed: LTT 3218 and EG 274 \citep{Hamuy1992}, for flux calibration purposes.

We are also presenting data from the MUSE \citep{Bacon2010} instrument attached to the ESO-VLT for Tol 1004-296. The observation was part of proposal ID 094.B-0745 (PI - Garcia-Benito, R.). The MUSE instrument has the great advantage of having a large Field of View (FoV) equal to $\approx $ 1.0 \arcmin $\times$ 1.0 \arcmin, allowing the observation of the entire object in a single 2880s observation. The spatial sampling is 0.2\arcsec~ per pixel. On the other hand, the available grating only provides a spectral resolving power of 1770@4800\AA.

\subsection{Data reduction}
\subsubsection{Preliminary Reduction}
The initial stage of data reduction, comprising bias correction, flat field correction, identification of lenses, and cube building, was conducted at the Laboratório Nacional de Astrofísica - LNA office. Wavelength calibration was also performed at LNA. The office sent us individual data cubes wavelength calibrated for each field for the two galaxies.

The first step was to construct a mosaic consisting of the four observed fields for each object.

Flux calibration was performed using the two spectrophotometric stars, LTT 3218 and EG 274, and the \textit{standard, sensfunc, calibrate} tasks from IRAF\footnote{IRAF - Image Reduction and Analysis Facility - \citet{Tody1986}}.

\subsection{Moment Maps}
The subsequent step involved extracting sub-cubes for each of the targeted emission lines. Depending on the lines, the data cube depth had between 30 or 40 spectral pixels, sufficient to determine the continuum level. To account for seeing effects, we applied a light spatial Gaussian smoothing of 3 spaxels (equivalent to 0.9\arcsec) Full Width at Half Maximum (FWHM).

Next, we generated various moment maps using two different techniques: an in-house macro and a Gaussian profile fitting technique.

\begin{itemize}

\item Monochromatic and Continuum Maps:
For each sub-cube, we identified the peak emission and isolated the profile. Then, we performed a linear fit of the remaining continuum and removed it from the sub-cube. Subsequently, we collapsed it to obtain monochromatic maps in different emission lines. We also generated continuum maps by removing the prominent emission lines and collapsing the cube, particularly for the MUSE data where the quality was better.

\item Radial Velocity Maps:
To produce the velocity map, we isolated the profile as described before and estimated the barycenter \footnote{It is estimated from histogram of line width, with the weight being the area of each rectangle forming the profile. More details in  \citet{Amram1991}} of the profile to account for possible asymmetries.

\item Velocity Dispersion Maps:
The velocity dispersion map was derived from the FWHM of the isolated profile. The FWHM was then transformed into dispersion using the formula 
$\sigma = \frac{FWHM}{2\sqrt{2ln2}} $. \\ 
The  velocity dispersion was corrected for thermal\footnote{See \citet{Rybicki2004}} and instrumental broadening following the formula.

$$\sigma = \sqrt{\sigma_{obs}^2 - \sigma_{th}^2 - \sigma_{inst}^2}   $$

The thermal dispersion, $\sigma_{th}$, was estimated to be 9.32 km s$^{-1}$ for hydrogen at a temperature of $10^4$ K. We estimated the instrumental broadening using the calibration lamp observation, giving 
$\sigma_{inst}$ = 13.0 $\pm$ 3 km s$^{-1}$.

\end{itemize}

Additionally, we used a Gaussian fit macro to cross-verify our results. However, we opted to retain the technique described above for estimating the velocity maps and velocity dispersion, as it allows for the consideration of possible asymmetries in the profiles.

\section{Results}     

\begin{table*}
	\centering
	\scriptsize
     	\caption[Intensity ratios (corrected from extinction), normalized to H$\beta$, interstellar extinction, diagnostic line ratios, densities, oxygen abundances, radial velocity and velocity dispersion, for the emission knots]{Intensity ratios (corrected from extinction), normalized to H$\beta$, interstellar extinction, diagnostic line ratios, densities, oxygen abundances, radial velocity and velocity dispersion, for the emission knots.\label{tab:fluxesTol1004296}}
	\begin{tabular}{lccccccc}
\hline\hline
\textbf{Ion} & \textbf{Knot 1}  & \textbf{Knot 2} & \textbf{Knot 3} & \textbf{Knot 4}  & \textbf{Knot 5} & \textbf{Knot 6}  & \textbf{Knot 7} \\
\hline
4861 H$\beta$ & 1.00  &  1.00  &  1.00  &  1.00  &  1.00  &  1.00  &  1.00  \\
4959 [OIII] & 1.70 $\pm$ 0.01 &  1.42 $\pm$ 0.01 &  1.51 $\pm$ 0.04 &  1.07 $\pm$ 0.08 &  1.30 $\pm$ 0.02 &  1.57 $\pm$ 0.05 &  1.11 $\pm$ 0.02  \\
5007 [OIII] & 5.13 $\pm$ 0.04 &  4.22 $\pm$ 0.03 &  4.70 $\pm$ 0.12 &  2.96 $\pm$ 0.23 &  3.75 $\pm$ 0.05 &  4.37 $\pm$ 0.14 &  3.40 $\pm$ 0.06  \\
6548 [NII] &0.04 $\pm$ 0.01 &  0.04 $\pm$ 0.01 &  0.02 $\pm$ 0.01 &  0.09 $\pm$ 0.01 &  0.03 $\pm$ 0.01 &  0.03 $\pm$ 0.02 &  0.03 $\pm$ 0.01  \\
6563 H$\alpha$  & 2.51 $\pm$ 0.02 &  2.51 $\pm$ 0.02 &  2.51 $\pm$ 0.06 &  2.51 $\pm$ 0.16 &  2.51 $\pm$ 0.03 &  2.51 $\pm$ 0.07 &  2.51 $\pm$ 0.04  \\
6583 [NII] &0.13 $\pm$ 0.01 &  0.16 $\pm$ 0.01 &  0.12 $\pm$ 0.01 &  0.13 $\pm$ 0.01 &  0.13 $\pm$ 0.01 &  0.12 $\pm$ 0.02 &  0.21 $\pm$ 0.02  \\
6713 [SII]  & 0.12 $\pm$ 0.01 &  0.17 $\pm$ 0.01 &  0.17 $\pm$ 0.02 &  0.24 $\pm$ 0.02 &  0.16 $\pm$ 0.01 &  0.10 $\pm$ 0.01 &  0.23 $\pm$ 0.01  \\
6732 [SII] & 0.10 $\pm$ 0.01 &  0.13 $\pm$ 0.01 &  0.12 $\pm$ 0.02 &  0.18 $\pm$ 0.01 &  0.13 $\pm$ 0.02 &  0.07 $\pm$ 0.01 &  0.17 $\pm$ 0.02  \\
\hline
F(H$\beta$)$^1$ & 99.69 $\pm$ 0.37 &  55.71 $\pm$ 0.18 &  3.50 $\pm$ 0.04 &  0.87 $\pm$ 0.03 &  2.86 $\pm$ 0.02 &  0.87 $\pm$ 0.01 &  1.99 $\pm$ 0.02  \\
EW(H$\beta$)$^2$ & 68.31 $\pm$ 0.12 &  76.32 $\pm$ 0.13 &  62.06 $\pm$ 0.40 &  48.48 $\pm$ 0.75 &  63.20 $\pm$ 0.24 &  61.32 $\pm$ 0.70 &  61.32 $\pm$ 0.7  \\
E(B-V) & 0.34 $\pm$ 0.03 &  0.313 $\pm$ 0.017 &  0.304 $\pm$ 0.037 &  0.178 $\pm$ 0.078 &  0.270 $\pm$ 0.030 &  0.209 $\pm$ 0.052 &  0.333 $\pm$ 0.044  \\
\hline 
\textbf{Parameters} & \textbf{Knot 1}  & \textbf{Knot 2} & \textbf{Knot 3} & \textbf{Knot 4}  & \textbf{Knot 5} & \textbf{Knot 6}  & \textbf{Knot 7} \\
log([OIII]5007/H$\beta$) & 0.65 $\pm$ 0.07 &  0.59 $\pm$ 0.04 &  0.65 $\pm$ 0.05 &  0.48 $\pm$ 0.08 &  0.57 $\pm$ 0.02 &  0.63 $\pm$ 0.02 &  0.52 $\pm$ 0.03  \\
log([NII]6583/H$\alpha$)& -1.16 $\pm$ 0.07 & -1.21 $\pm$ 0.13 & -1.33 $\pm$ 0.13 & -1.27 $\pm$ 0.17 &  -1.23 $\pm$ 0.07 &  -1.27 $\pm$ 0.08 & -1.01 $\pm$ 0.09  \\
log([SII]6716 6732/H$\alpha$)&-0.93 $\pm$ 0.17 &  -0.86 $\pm$ 0.11 & -0.93 $\pm$ 0.10 &  -0.80 $\pm$ 0.21 & -0.90 $\pm$ 0.05 & -1.17 $\pm$ 0.16 & -0.77 $\pm$ 0.07  \\
\hline
n$_e$ (cm$^{-3}$)  & < 100   & < 100 & < 100  & < 100 & < 100 & < 100 & < 100  \\
12+log(O/H) (S calibrator)$^3$  & 8.14 $\pm$ 0.04 &  8.14 $\pm$ 0.02 &  8.10 $\pm$ 0.06 &  8.14 $\pm$ 0.23 &  8.07 $\pm$ 0.04 &  8.12 $\pm$ 0.09 &  8.26 $\pm$ 0.12  \\
12+log(O/H) (O3N2 Calibrator)$^4$  & 8.06 $\pm$ 0.02 &  8.12 $\pm$ 0.02 &  8.10 $\pm$ 0.03 &  8.15 $\pm$ 0.04 &  8.14 $\pm$ 0.01 &  8.12 $\pm$ 0.02 &  8.19 $\pm$ 0.01  \\
12+log(O/H) (N2 Calibrator)$^5$  & 8.08 $\pm$ 0.03 &  8.16 $\pm$ 0.03 &  8.16 $\pm$ 0.03 &  8.12 $\pm$ 0.03 &  8.16 $\pm$ 0.02 &  8.14 $\pm$ 0.05 &  8.26 $\pm$ 0.03  \\
\hline
\textbf{Kinematics} & \textbf{Knot 1}  & \textbf{Knot 2} & \textbf{Knot 3} & \textbf{Knot 4}  & \textbf{Knot 5} & \textbf{Knot 6}  & \textbf{Knot 7} \\
Radial Velocity (km s$^{-1}$) & 1137.0$\pm$ 5.0 &  1162.0 $\pm$ 4.0 &  1175.0 $\pm$ 5.0 &  1205.0 $\pm$ 6.0 &  1147.0 $\pm$ 4.0 &  1176.0 $\pm$ 6.0 &  1185.0 $\pm$ 7.0  \\
Velocity Dispersion (km s$^{-1}$) & 31.0 $\pm$ 2.0 &  34.0 $\pm$ 2.0 &  20.0 $\pm$ 2.0 &  17.0 $\pm$ 5.0 &  28.0 $\pm$ 2.0 &  20.0 $\pm$ 2.0 &  25.0 $\pm$ 2.0  \\

\hline
       \end{tabular}
       
\flushleft Notes: $^1$: H$\beta$ fluxes in 10$^{-14}$ erg s$^{-1}$ cm$^{-2}$. $^2$ In \AA. $^3$ S Calibrator Oxygen abundance from \citet{Pilyugin2016}. $^4$ O3N2 Calibrator from \citet{Marino2013}. $^5$ N2 Calibrator from \citet{Marino2013}.
\end{table*}       

%
%
%

\begin{table*}
	\centering
	\scriptsize
     	\caption[Intensity ratios (corrected from extinction), normalized to H$\beta$, interstellar extinction, diagnostic line ratios, densities, oxygen abundances, radial velocity and velocity dispersion, for the emission knots]{Intensity ratios (corrected from extinction), normalized to H$\beta$, interstellar extinction, diagnostic line ratios, densities, oxygen abundances, radial velocity and velocity dispersion, for the emission knots.\label{tab:fluxesTol0957278}}
	\begin{tabular}{lccccccccc}
\hline\hline
\textbf{Ion} & \textbf{Knot 1}  & \textbf{Knot 2} & \textbf{Knot 3} & \textbf{Knot 4}  & \textbf{Knot 5} & \textbf{Knot 6}  & \textbf{Knot 7}  & \textbf{Knot 8}  & \textbf{Knot 9} \\
\hline
4861 H$\beta$ & 1.00  &  1.00  &  1.00  &  1.00  &  1.00  &  1.00  &  1.00  &  1.00  & 1.00  \\
4959 [OIII] & 1.38 $\pm$ 0.02 &  2.00 $\pm$ 0.04 &  1.88 $\pm$ 0.04 &  1.44 $\pm$ 0.01 &  1.44 $\pm$ 0.02 &  1.42 $\pm$ 0.02 &  1.43 $\pm$ 0.02 &  1.64 $\pm$ 0.04 &  1.73 $\pm$ 0.06  \\
5007 [OIII] & 4.31 $\pm$ 0.05 &  6.47 $\pm$ 0.13 &  5.73 $\pm$ 0.13 &  4.44 $\pm$ 0.04 &  4.28 $\pm$ 0.07 &  4.53 $\pm$ 0.06 &  4.48 $\pm$ 0.05 &  4.82 $\pm$ 0.10 &  5.21 $\pm$ 0.19  \\
6563 H$\alpha$  & 3.06 $\pm$ 0.03 &  3.06 $\pm$ 0.05 &  3.06 $\pm$ 0.06 &  3.06 $\pm$ 0.03 &  3.06 $\pm$ 0.04 &  3.06 $\pm$ 0.03 &  3.06 $\pm$ 0.03 &  3.06 $\pm$ 0.06 &  3.06 $\pm$ 0.10  \\
6583 [NII] & 0.12 $\pm$ 0.01 &  0.46 $\pm$ 0.01 &  0.06 $\pm$ 0.01 &  0.12 $\pm$ 0.01 &  0.10 $\pm$ 0.02 &  0.09 $\pm$ 0.01 &  0.12 $\pm$ 0.02 &  0.09 $\pm$ 0.01 &  0.15 $\pm$ 0.01  \\
6713 [SII]  & 0.21 $\pm$ 0.02 &  0.08 $\pm$ 0.01 &  0.18 $\pm$ 0.01 &  0.22 $\pm$ 0.02 &  0.19 $\pm$ 0.01 &  0.21 $\pm$ 0.01 &  0.28 $\pm$ 0.03 &  0.21 $\pm$ 0.02 &  0.16 $\pm$ 0.02  \\
6732 [SII] & 0.14 $\pm$ 0.01 &  0.08 $\pm$ 0.02 &  0.08 $\pm$ 0.01 &  0.14 $\pm$ 0.01 &  0.11 $\pm$ 0.02 &  0.15 $\pm$ 0.01 &  0.20 $\pm$ 0.02 &  0.16 $\pm$ 0.02 &  0.10 $\pm$ 0.01  \\
\hline
F(H$\beta$)$^1$ &  18.77 $\pm$ 0.11 &  4.81 $\pm$ 0.04 &  2.89 $\pm$ 0.03 &  4.21 $\pm$ 0.02 &  1.23 $\pm$ 0.01 &  2.37 $\pm$ 0.01 &  3.08 $\pm$ 0.02 &  1.69 $\pm$ 0.02 &  0.82 $\pm$ 0.01  \\
EW(H$\beta$)$^2$ &  65.16 $\pm$ 0.06 &  67.54 $\pm$ 0.07 &  34.96 $\pm$ 0.13 &  39.95 $\pm$ 0.03 &  34.11 $\pm$ 0.04 &  34.28 $\pm$ 0.06 &  32.28 $\pm$ 0.06 &  32.813 $\pm$ 0.07 &  34.19 $\pm$ 0.07  \\
E(B-V)  & 0.511 $\pm$ 0.10 &  0.33 $\pm$ 0.03 &  0.48 $\pm$ 0.04 &  0.36 $\pm$ 0.03 &  0.33 $\pm$ 0.03 &  0.41 $\pm$ 0.04 &  0.41 $\pm$ 0.05 &  0.45 $\pm$ 0.07 &  0.34 $\pm$ 0.04  \\
\hline 
\textbf{Parameters} & \textbf{Knot 1}  & \textbf{Knot 2} & \textbf{Knot 3} & \textbf{Knot 4}  & \textbf{Knot 5} & \textbf{Knot 6}  & \textbf{Knot 7} & \textbf{Knot 8}  & \textbf{Knot 9} \\
log([OIII]5007/H$\beta$) & 0.63 $\pm$ 0.05 &  0.78 $\pm$ 0.07 &  0.75 $\pm$ 0.03 &  0.64 $\pm$ 0.04 &  0.63 $\pm$ 0.02 &  0.75 $\pm$ 0.03 &  0.64 $\pm$ 0.04 &  0.68 $\pm$ 0.03 &  0.71 $\pm$ 0.02  \\
log([NII]6583/H$\alpha$) & -1.39 $\pm$ 0.09 &  -1.45 $\pm$ 0.09 &  -1.47 $\pm$ 0.09 &  -1.39 $\pm$ 0.07 &  -1.45 $\pm$ 0.07 &  -1.44 $\pm$ 0.10 &  -1.37 $\pm$ 0.11 &  -1.41 $\pm$ 0.08 &  -1.31 $\pm$ 0.16  \\
log([SII]6716 6732/H$\alpha$) & -0.93 $\pm$ 0.06 &  -1.22 $\pm$ 0.13 &  -1.14 $\pm$ 0.09 &  -0.94 $\pm$ 0.07 &  -0.10 $\pm$ 0.09 &  -0.92 $\pm$ 0.07 &  -0.80 $\pm$ 0.07 &  -0.95 $\pm$ 0.12 & -1.07 $\pm$ 0.08  \\
\hline
n$_e$ (cm$^{-3}$)  & < 100   & < 100 & < 100  & < 100 & < 100 & < 100 & < 100 & < 100 & < 100 \\
12+log(O/H) (N2 Calibrator)$^3$ & 8.09 $\pm$ 0.02 &  7.84 $\pm$ 0.16 &  8.00 $\pm$ 0.13 &  8.11 $\pm$ 0.04 &  8.07 $\pm$ 0.04 & 8.05 $\pm$ 0.11 &  8.09 $\pm$ 0.12 &  7.95 $\pm$ 0.21 & 8.14 $\pm$ 0.08  \\
\hline
\textbf{Kinematics} & \textbf{Knot 1}  & \textbf{Knot 2} & \textbf{Knot 3} & \textbf{Knot 4}  & \textbf{Knot 5} & \textbf{Knot 6}  & \textbf{Knot 7} & \textbf{Knot 8}  & \textbf{Knot 9} \\
Radial Velocity (km s$^{-1}$) &  1041.0 $\pm$ 5.0 &  1016.0$\pm$ 4.0 &  1027.0 $\pm$ 5.0 &  1030.4 $\pm$ 34.0&  1032.0 $\pm$ 3.0 &  1029.0 $\pm$ 5.0 &  1030.0 $\pm$ 4.0 &  1029.0 $\pm$ 5.0 &  1025.0 $\pm$ 4.0  \\
Velocity Dispersion  (km s$^{-1}$) &  21.0 $\pm$ 3.0 &  20.0 $\pm$ 3.0 &  22.0 $\pm$ 2.0 &  27.0 $\pm$ 3.0&  20.0 $\pm$ 2.0 &  23.5 $\pm$ 3.0 &  24.0 $\pm$ 2.0 &  26.0 $\pm$ 3.0 &  24.0 $\pm$ 3.0  \\
\hline
       \end{tabular}
       
\flushleft Notes: $^1$: H$\beta$ fluxes in 10$^{-14}$ erg s$^{-1}$ cm$^{-2}$. In \AA. $^3$  N2 Calibrator from \citet{Marino2013}.
\end{table*}       

\subsection{Emission lines maps}
\begin{figure*}
	\includegraphics[width=\textwidth]{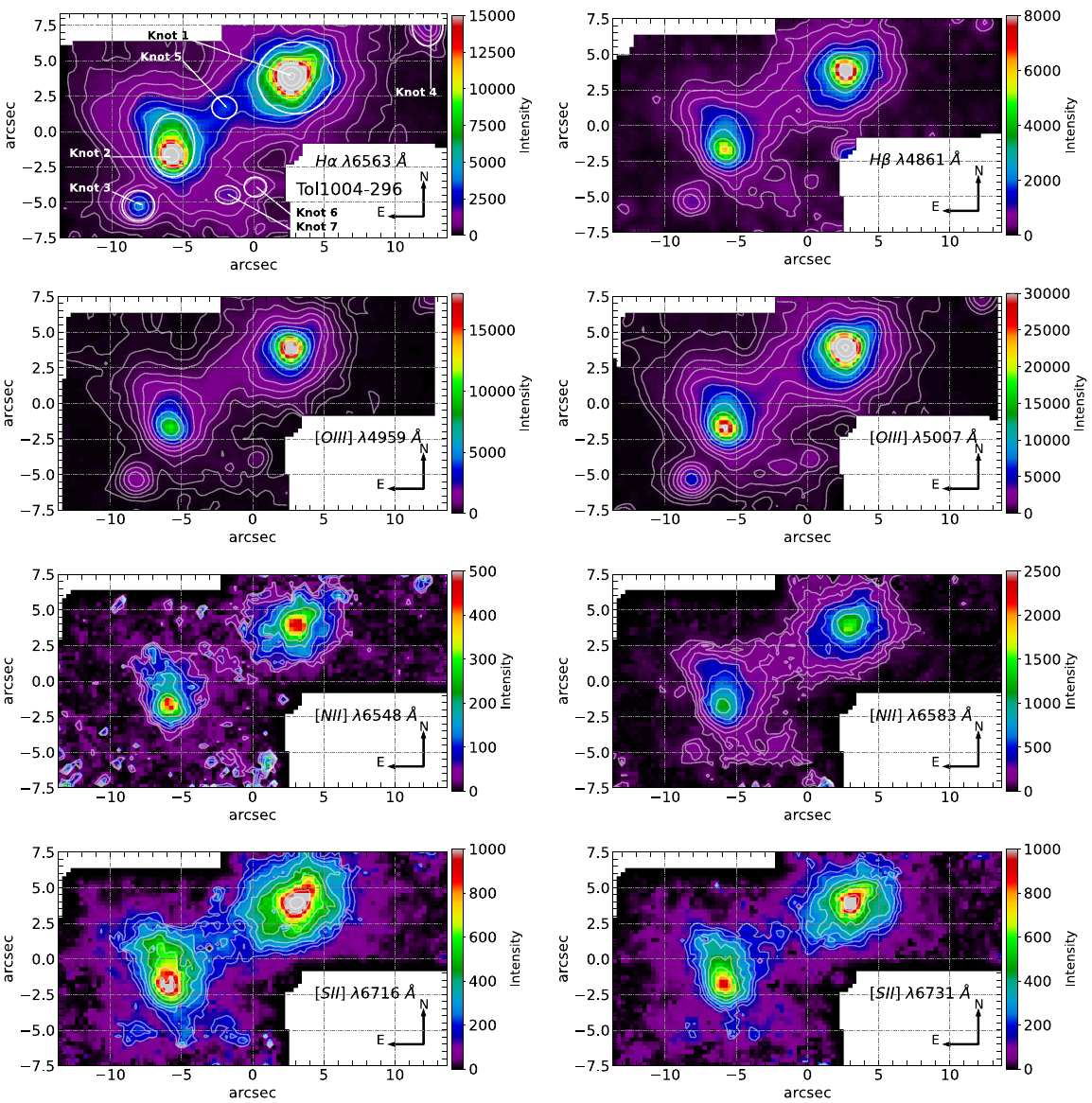}
    \caption{Monochromatic maps for Tol 1004-296.} 
    \label{fig:MonoMapsITol1004-296}
\end{figure*}

\begin{figure*}
	\includegraphics[width=\textwidth]{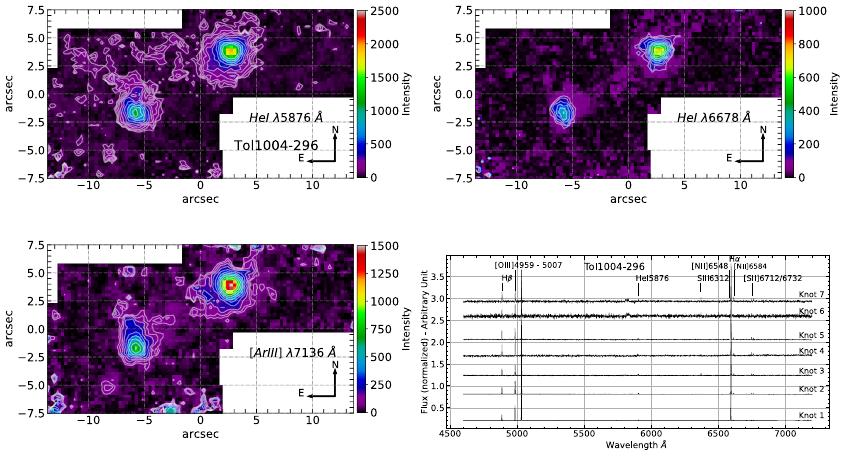}
    \caption{Monochromatic maps for Tol 1004-296 and integrated spectra of the seven emission knots. Each spectrum represent a 3$\times$3 integrated spaxels.} 
    \label{fig:MonoMapsIITol1004-296}
\end{figure*}

\begin{figure*}[h]
	\includegraphics[width=\textwidth]{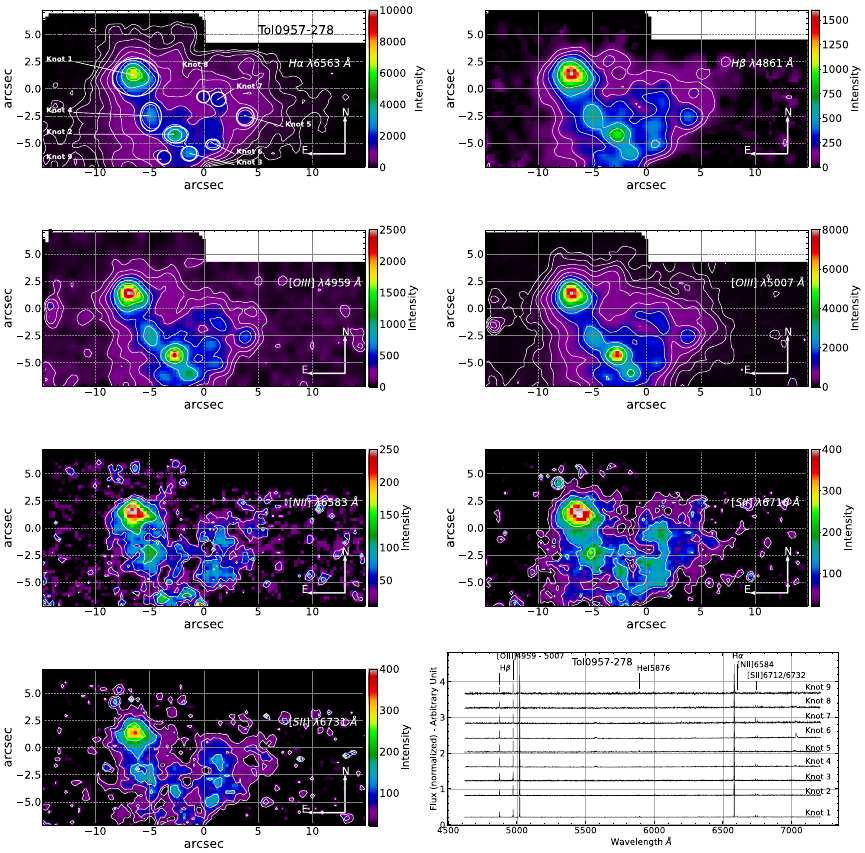}
    \caption{Monochromatic maps for Tol 0957-278 and integrated spectra of the nine emission knots. Each spectrum represent a 3$\times$3 integrated spaxels.} 
    \label{fig:MonoMapsITol0957-278}
\end{figure*}

We were able to produce monochromatic maps of the following emission lines: H$\beta\lambda$4861, [OIII]$\lambda\lambda$4959, 5007, HeI$\lambda$5876, [NII]$\lambda$6548, H$\alpha\lambda$6563, [NII]$\lambda$6584, [SII]$\lambda\lambda$6716, 6732, HeI$\lambda$6678, [ArIII]$\lambda$7136. Figure \ref{fig:MonoMapsITol1004-296} to Figure \ref{fig:MonoMapsITol0957-278} present these different monochromatic maps for both objects (except that for Tol 0957-278, where [HeI]$\lambda$5876, [NII]$\lambda$6548, [HeI]$\lambda$6678 and [ArIII]$\lambda$7136 lines were too weak to be extracted). Table \ref{tab:fluxesTol1004296} and Table  \ref{tab:fluxesTol0957278} summarize the fluxes of the different emission lines in the different emission knots (they also include several others quantities such as diagnostic line ratios, oxygen abundances, interstellar extinction, radial velocity and velocity dispersion). These knots are symbolically shown in Figure \ref{fig:MonoMapsITol1004-296} (H$\alpha$ map) and Figure \ref{fig:MonoMapsITol0957-278} (H$\alpha$ map). We estimated these regions by limiting the extension where the SNR is above 5 in the $H\alpha$ line. In Figure \ref{fig:Histograms}, we have added several histograms representing the distribution of several parameters ([SII]$\lambda$6716/[SII]$\lambda$6731, EW(H$\beta$), E(B-V) and diagnostics diagrams line ratios). \\

In Tol 1004-296 case, we recognised the two large knots (Knot 1 and 2) separated by 10.3\arcsec~(661 pc), visible in Figure \ref{fig:LTO_Tol1004_Tol0957}.

It is noticeable that only the two largest bright sites appear weaker in others emission wavelengths, such as [NII]$\lambda$6584, [SII]$\lambda$6716, [SII]$\lambda$6716, [HeI]$\lambda$5876, [NII]$\lambda$6548, [HeI]$\lambda$6678 and [ArIII]$\lambda$7136. We can also notice that the Knot 7 has a more elongated shape or formed by two sub nuclei in [OIII]$\lambda$5007 or H$\alpha$.

The monochromatic maps from the MUSE data are available in Figures~\ref{fig:MonoMapsTol1004-296MUSE}. Extension of the ionized gas is, as expected, much larger compared to our SIFS data. For the  brightest emission lines (H$\alpha$, H$\beta$, [OIII]$\lambda\lambda$ 4959, 5007), the extension is 50\arcsec $\times$ 50\arcsec~ (we limit the emission to a Signal to Noise Ratio (SNR) superior to 8, in order to keep the filamentary character of the emission, in the outskirt, without the sky background noise). It is more or less the same extension showed by \citet{GildePaz2003} with the narrow band imaging of the H$\alpha$ + [NII] emission.
The MUSE data show the following emission lines: H$\beta~\lambda$4861, [OIII]$\lambda\lambda$4959, 5007, HeI$\lambda$5876, [OI]$\lambda$6300, [NII]$\lambda$6548, H$\alpha\lambda$6563, [NII]$\lambda$6584, [SII]$\lambda\lambda$6716, 6732, HeI$\lambda$6678, [ArIII]$\lambda$7136. We easily recognise the different Knots we have presented previously. \\

Tol 0957-278 maps show a more complex morphology (Figure \ref{fig:MonoMapsITol0957-278}), nine emission sites can be seen on the $[OIII]\lambda4959$ and $[OIII]\lambda5007$ lines. 
H$\alpha$ map in Figure ~\ref{fig:MonoMapsITol0957-278} shows the location of emission sites in the galaxy. Four knots (1, 2, 3 and 4) are brighter than the other five knots (5, 6, 7, 8 and 9).

\subsection{Electron density and electron temperature}
\subsubsection{Electron density}
\begin{figure*}
	\includegraphics[width=\textwidth]{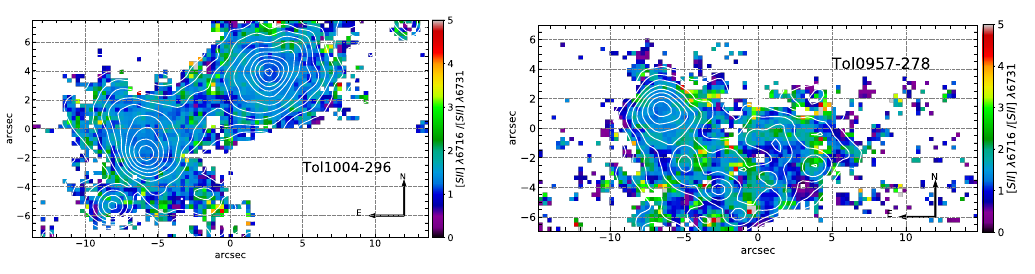}
    \caption{[SII]$\lambda$6717/[SII]$\lambda$6731 maps for Tol 1004-296 and Tol0957-278. For each galaxy, H$\alpha$ emission isocontours are presented} 
    \label{fig:Map[SII]6717_6731Tol1004-296_Tol0957-278}
\end{figure*}

In order to estimate the abundances, it is first necessary to estimate the electronic density and temperature. The electron density is obtained using the emission line ratio [SII]$\lambda$6717 / [SII]$\lambda$6731 and using the task TEMDEM, based on the FIVEL program \citep{Shaw1995}, which is included in the IRAF package NEBULAR. We also used the PyNeb code \citet{Luridiana2015}.
[SII]$\lambda$6717 / [SII]$\lambda$ 6731 maps are presented in Figure \ref{fig:Map[SII]6717_6731Tol1004-296_Tol0957-278} for both objects. The ratio is between 0.7  $\pm$ 0.1 and 1.8 $\pm$ 0.1 or Tol 1004-296 (some hot pixels have a value up to 3.5) and between 0.65  $\pm$ 0.1 and 1.95 $\pm$ 0.1 for Tol 0957-278 (here too, some hot pixels have values up to 3.5).
The center of Knots 1 and 2 of Tol 1004-296 have lower values compared to values around. In Tol 0957-278 case too, emission knots show lower ratio compared to values around.
This gives an electron density 80 cm$^{-3}$ < $n_e$ < 400 cm$^{-3}$. Following the low regime \citet{Osterbrock2006}, we assume the electron density $n_e$ = 100 cm$^{-3}$ for both galaxies \citep{Lagos2009, GarciaLorenzo2008, Cairos2020, Cairos2022}.

\subsubsection{Electron temperature}
The electron temperature, $T_e$ is usually estimated using the oxygen emission line, with [OIII]$\lambda\lambda$4959, 5007 / [OIII]$\lambda$4363 ratio and one of the code mentioned above (NEBULAR or Pyneb). The [OIII]$\lambda$4363 is outside our spectral range. We decided to use the values of these lines available in \citet{Kehrig2004}. For both galaxies, \citet{Kehrig2004} is giving three values (center, NW and SE), we then used PyNeb to estimate the electron temperature, assuming a $n_e$ = 100 cm$^{-3}$. Table \ref{tab:electrontemp} summarises the different values for both objects.

\begin{table}
	\caption[Electron Temperature]{Electron Temperature \label{tab:electrontemp}}
	\begin{tabular}{lcc}
\textbf{Name} & \textbf{[OIII]$\lambda\lambda$4959, 5007 / [OIII]$\lambda$4363} & \textbf{$T_e$ (K)}   \\
\hline\hline
Tol1004-296          & 180.93 &  10511  \\
Tol1004-296(NW)  & 199.56 &  10193 \\
Tol1004-296(SE)   &170.36  &  10716  \\
                              &                                                                                                      &       \\
Tol0957-278          & 107.87 & 12580 \\
Tol0957-278(NW) & 140.34 &  11435 \\
Tol0957-278(SE)  & 73.730 &  14710 \\
\hline
	\end{tabular}
\end{table}

PyNeb \citep{Luridiana2015} also allowed us to diagnose the electron density and temperature. With the values of the [OIII]$\lambda\lambda$ 4959, 5007 / [OIII]$\lambda$ 4363 and [SII]$\lambda$6717 / [SII]$\lambda$ 6731 ratios from \citet{Kehrig2004}, it was possible to have an $T_e$ vs $n_e$ graph with the [OIII] lines and [SII] ratios. This graph is compatible with the use of a constant $n_e$ = 100 cm$^{-3}$ and $T_e$ = 10$^4$ ~K.

Figure \ref{fig:LinesDisgnosticVelocityMapsTol1004-296MUSE} shows the [SII]$\lambda$6716/[SII]$\lambda$6732 lines ratio for Tol 1004-296 observed with MUSE. In the center, corresponding to the SIFS FoV, Knot 1 and Knot 2 show respectively 1.25  $\pm$ 0.07 and 1.29 $\pm$ 0.04.

the ratio is between 1.15 and 1.4. The minimum ratio corresponds to the two principals emission knots (Knots 1 and 2).

\subsection{Interstellar extinction}
Dust extinction has to be estimated and then applied to the observed fluxes in order to have correct fluxes.
Interstellar extinction in nebula is computed using the observed Balmer decrement.
The intrinsic luminosity, $L_{int}$ in a given wavelength can be written as:

$$L_{int}(\lambda)= L_{obs}(\lambda) ~10^{0.4k(\lambda)E(B-V)}$$

with $L_{obs}(\lambda)$ the observed luminosity, $k(\lambda)$ the extinction law and $E(B-V)$ the color excess \citep{Calzetti1994, Dominguez2013}. We used the \citet{Calzetti2000} extinction law for $k(\lambda)$ and the color excess $E(B-V)$ has been derived from the Balmer decrement using $H\alpha$ and $H\beta$ lines by:

$$E(B-V) = \frac{E(H\beta - H\alpha)}{k(\lambda_{H\beta}) - k(\lambda_{H\alpha})} $$ 
$$                = \frac{2.5}{k(\lambda_{H\beta}) - k(\lambda_{H\alpha})} ~ \log_{10} \left [ \frac{(H\alpha/H\beta)_{obs}}{(H\alpha/H\beta)_{int}} \right ] $$ 

with $k(\lambda_{H\beta})$ and $k(\lambda_{H\alpha})$ the extinction coefficients for $H\beta$ and $H\alpha$.

Adopting the theoretical ratio of 2.86  (case B recombination at a temperature of $10^4$ K; see \citet{Osterbrock2006}), the color excess is:

$$E(B-V) =  1.97~\log_{10} ~ \left [ \left (\frac{H\alpha}{H\beta} \right )_{obs}.\frac{1}{2.86} \right ] $$ 
 
\begin{figure*}
	\includegraphics[width=\textwidth]{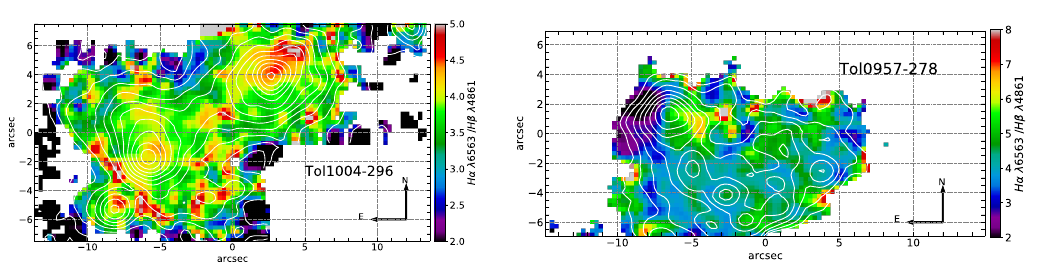}
    \caption{$H\alpha$/$H\beta$ maps for Tol 1004-296 and Tol0957-278. For each galaxy, H$\alpha$ emission isocontours are presented.} 
    \label{fig:Halpha_HbetaTol1004-296_Tol0957-278}
\end{figure*}

Figure \ref{fig:Halpha_HbetaTol1004-296_Tol0957-278} shows $H\alpha$/$H\beta$ ratio maps for both objects. For Tol 1004-296, the ratio is more or less uniform around 3.8 $\pm$ 0.4 with some fluctuation. Some hot spots are visible where the emission is higher (where the strong emission sites are), as for the Knot 1 (where the ratio is 4.3 $\pm$ 0.2), the Knot 2  (where the ratio is 4.2 $\pm$ 0.1) and the Knot 3 (where the ratio is 4.0 $\pm$ 0.3).

For Tol 0957-278, the average ratio is around 4.3 $\pm$ 0.9. It is notable that the Knot 1 shows a value of 5.7 $\pm$ 0.5 and 3.4\arcsec East, the ratio goes down to 2 $\pm$ 0.4. It is possible that this difference is coming from a dust lane, present in this area.

Figure \ref{fig:LinesDisgnosticVelocityMapsTol1004-296MUSE} shows the $H\alpha / H\beta$ line ratio for Tol 1004-296 observed with MUSE. The line ratio inside the SIFS FoV is consistent with the map from the SIFS data.

\subsection{Chemical abundances}

It wasn't possible to use the direct T$_e$ method \citep[e.g.][]{Dinerstein1990, McCall1985}, to determine the oxygen abundance because 1] the auroral lines such as [OIII]$\lambda$4363 and [NII]$\lambda$5755 were too weak to be detected and 2] the [OII]$\lambda\lambda$3727, 3729 doublet and [OII]$\lambda\lambda$7320, 7331 lines are outside of the SIFS spectral range.
However, several studies \citep[e.g.][]{Pilyugin2012, Pilyugin2016, Marino2013, Denicolo2002} offer a series of calibrators in order to estimate the oxygen abundance using a variety of available emission lines and based on statistical studies.

\citet{Denicolo2002} and \citet{Marino2013} give the so-called $N2$ calibrator based on the [NII]$\lambda$6584 / H$\alpha$ line ratio, with slightly different regression coefficients. First introduced by \citet{Alloin1979}, \citet{Marino2013} detailed the so-called $O3N2$ calibrator based on the [OIII]$\lambda$5007 / H$\beta$ $\times$ H$\alpha$ / [NII]$\lambda$6584 using the CALIFA survey. Introduced by \citet{Pilyugin2016}, the so-called $S$ calibrator used the ratios $([NII]6548 + 6583) / H\beta$, $([SII]6717 + 6732) / H\beta$ and $([OIII]4959 + 5007) / H\beta$ to derive the oxygen abundance. It's important to note from its definition that the $S$ calibrator depends on the reddening correction.

\begin{figure*}
	\includegraphics[width=\textwidth]{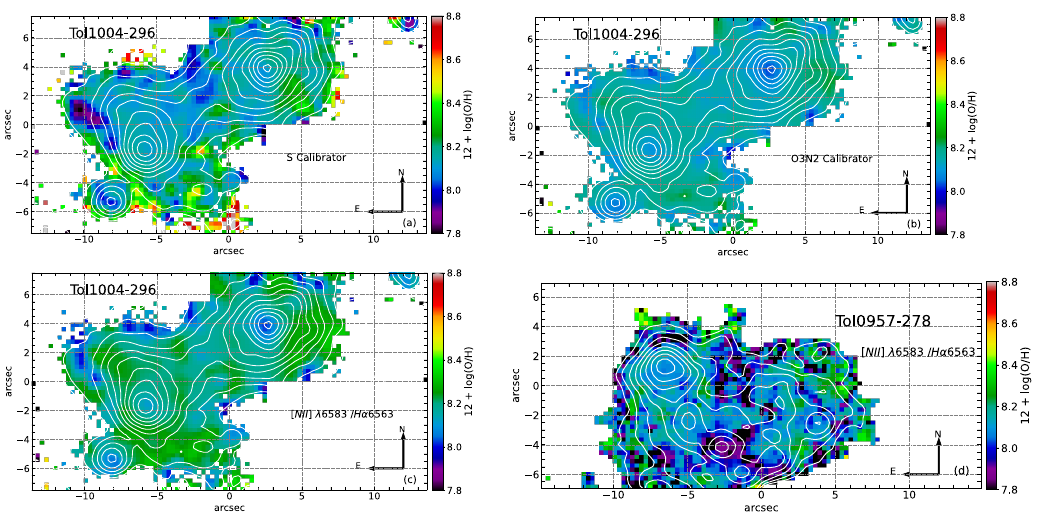}
    \caption{Metalicitty for Tol 1004-296 and Tol0957-278. For each galaxy, H$\alpha$ emission isocontours are presented. (a): Tol 1004-296 Metallicity map using the $S calibrator$, (b): Tol 1004-296 Metallicity map using the O3N2 calibrator, (c):  Tol 1004-296 Metallicity map using [NII]$\lambda$6583 / H$\alpha$ calibrator,  (d): Tol 0957-296 Metalicity map using the [NII]$\lambda$6583 / H$\alpha$ calibrator} 
    \label{fig:MetallicityTol1004-296_Tol0957-278}
\end{figure*}

For Tol 1004-296, we were able to implement the three methods. Figures \ref{fig:MetallicityTol1004-296_Tol0957-278}a, b, and c are respectively using the $S$ calibrator, the $O3N2$ calibrator, and $N2$ calibrator. Even if the overall aspect of the maps are similar, showing gradient between the different knots, we can notice some differences between the different methods. The $S$ calibrator (Fig. \ref{fig:MetallicityTol1004-296_Tol0957-278}a) shows the greater contrast in the map between different parts of the galaxy. Even if it's not the minimum of metallicity, the Knots 1, 2, and 3 show respectively a metallicity of 8.14  $\pm$ 0.04 dex, 8.14 $\pm$ 0.02 dex, and  8.10 $\pm$ 0.06 dex. On the outskirts (at 2.8 \arcsec from the Knots centers) the metallicity is higher, from 8.20 $\pm$ 0.05 dex, to 8.55  $\pm$ 0.05 dex. On the other hand, the $O3N2$ calibrator (Fig. \ref{fig:MetallicityTol1004-296_Tol0957-278}b) gives a lower amplitude between the metallicity in the emission knots and around them (the highest metallicity values do not exceed 8.22 $\pm$ 0.04 dex,). Finally, the $N2$ calibrator shows, in a clearer way, that higher metallicity is surrounding the emission knots.

We can notice that the metallicity of the different emission knots is between 8.06 $\pm$ 0.03 dex, and 8.16 $\pm$ 0.04 dex, except in Knot 7 (where it is 8.26 $\pm$ 0.3 dex).  It is worth notice that a lower metallicity seems to be present where the emission is higher. This is more noticeable for the Knot 1 and 2 and when looking at the calibrators O3N2 and N2.

Figure \ref{fig:LinesDisgnosticVelocityMapsTol1004-296MUSE} shows the metallicity map (using the S calibrator) for the MUSE data of Tol 1004-296. It can be noticed that the metallicity is slightly higher on the outskirts of the galaxy. In the center of the galaxy (essentially the field of view of the SIFS data), the metallicity ranges between 8.07 $\pm$ 0.04 dex and 8.30 $\pm$ 0.04 dex. As mentioned before, we also observed that lower metallicity corresponds to high emission knots. All seven emission knots (see Fig. ~\ref{fig:MonoMapsITol1004-296}) correspond to a minimum metallicity value.

In Tol 0957-278 case, we have implemented the N2 and O3N2 calibrators from \citet{Marino2013} (because the [NII]$\lambda$ 6548 line was too weak to be detected).
Figure \ref{fig:MetallicityTol1004-296_Tol0957-278}d shows the metallicity deduced from the $N2$ calibrator. It is clear here that the metallicity is lower where the emission is higher. The lowest metallicity is 7.84 $\pm$ 0.16 dex in Knot 2. The brightest site (Knot 1) has a metallicity of 8.09 $\pm$ 0.02 dex. Around Knot 1 (at a radius of 1.8 \arcsec from the brightest spot), the metallicity ranges between 8.18 $\pm$ 0.02 dex and 8.25 $\pm$ 0.03 dex.

\subsection{Ionization mechanism}
Photoionization from UV photons coming from hot stars, photoionization from an AGN source, or shocks from collisions (from stellar winds and supernovae \citep{Dopita2003}), are the three ionization mechanisms could be present in an emission-line system. \citet{Baldwin1981} and \citet{Veilleux1987} came up with a series of excitation-dependent line ratios, the so-called ionization diagnostic diagrams. The three diagnostic plots usually are [OIII]$\lambda$5007 / H$\beta$ versus [NII]$\lambda$6584 / H$\alpha$, [OIII]$\lambda$5007 / H$\beta$ versus [S II]$\lambda\lambda$6717, 6731 / H$\alpha$, and [OIII]$\lambda$5007 / H$\beta$ versus [O I]$\lambda$6300 / H$\alpha$. The way it works is to mix in the plot hard ionizing emission lines ratio ([OIII]$\lambda$5007 / H$\beta$) with low ionization lines ([NII]$\lambda$6584, [S II]$\lambda\lambda$6717, 6731) known to be present where hydrogen is partially ionized. High [NII]$\lambda$6584 / H$\alpha$, or [S II]$\lambda\lambda 6717, 6731$ / H$\alpha$ ratios are usually indicators that the ionization mechanism is more susceptible to be shock and/or AGN than photoionization from hot star UV photons. Then the three diagnostic diagrams are divided into three zones: photoionization from hot star UV photons (commonly noted \HII\ regions), AGN, and Shocks (noted as LINER). The diagnostic diagrams, commonly called BPT diagrams, from the name of the authors of \citet{Baldwin1981}, are presented in Figure \ref{fig:BPT_PlotMapTol1004-296}, Figure \ref{fig:BPT_PlotTol1004-296MUSE}, and Figure \ref{fig:BPT_PlotMapTol0957-278}. Except for the MUSE data, we were not able to detect the $[OI]\lambda 6300$ line, so for data coming from SIFS - SOAR, we only present the first two BPT diagrams.

\subsubsection{Tol 1004-296}
\begin{figure}
	\includegraphics[width=0.45\textwidth]{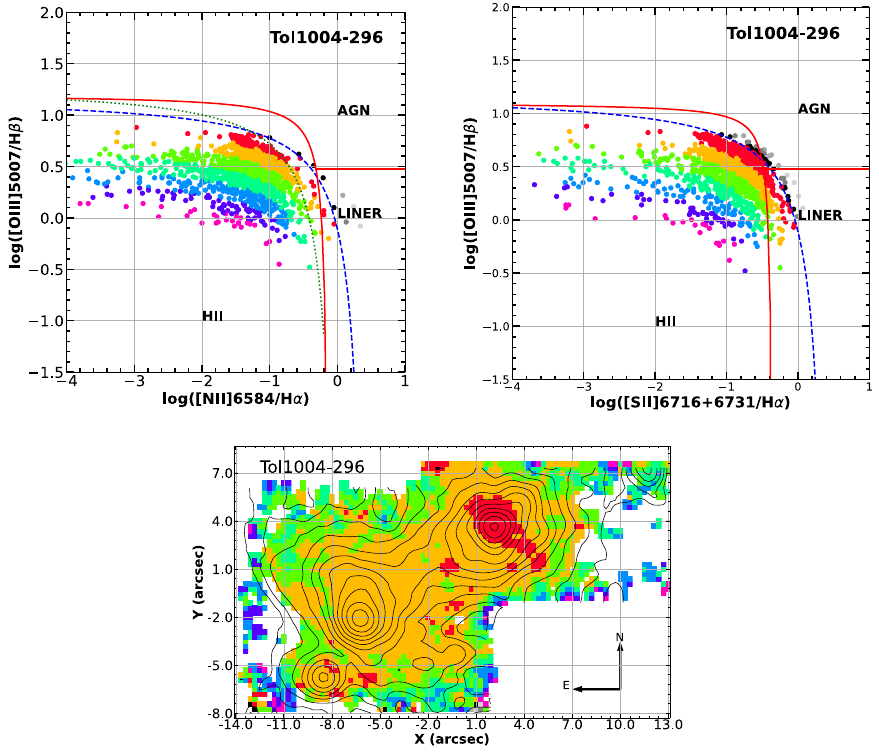}
    \caption{BPT diagrams. Color coded represent the distance from the \citet{Kewley2001} (blue dash line) separation. Also shown the \citet{Veilleux1987} (red lines) and \citet{Kauffman2003} (dotted line) limits $[OIII]5007/H\beta~vs~[NII]6584/H\alpha$ (Upper left). $[OIII]5007/H\beta~vs~[SII]6716+6731/H\alpha$ (Upper right). Spatial  locations of the spaxels from the $[OIII]5007/H\beta~vs~[NII]6584/H\alpha$ diagram. The color code is the same as the diagram above (Bottom).} 
    \label{fig:BPT_PlotMapTol1004-296}
\end{figure}

On Figure \ref{fig:BPT_PlotMapTol1004-296}(Top), we can see that, on both plots, as expected, almost all pixels are located on the photoionization by hot stars (\HII\ on the plot). The ratios ranges are $-4.0 < \log([\text{NII}] \lambda 6584 / H\alpha) < -0.5$, $-0.5 < \log([\text{OIII}] \lambda 5007 / H\beta) < -0.8$, and $-1.5 < \log([\text{S II}] \lambda\lambda 6717, 6731 / H\alpha) < -0.1$. Following previous work \citep{Cairos2017a}, we color-code the different points on the plot depending on the distance from the frontier between \HII\ and AGN regions. Both plots also exhibit the different frontiers from \citet{Veilleux1987, Kewley2001, Kauffman2003}. The diagram between $\log([\text{OIII}] \lambda 5007 / H\beta$ and  $\log([\text{S II}] \lambda\lambda 6717, 6731 / H\alpha$ shows that a few points are in the LINER area. The bottom map in Figure \ref{fig:BPT_PlotMapTol1004-296} shows the spatial distribution of the points in the diagrams (upper panels). The color coding is the same. Red to Violet represent the color of points in the \HII\ zone. The bluer the color, the further from the frontier the point is, from the maximum starburst line. The brightest knot (Knot 1) shows the largest concentration of red points, closer to the maximum starburst line in the diagnostic diagram. Then, around the Knot 1 and Knot 2, the points are orange and correspond to the points a bit further from the frontier.

Figures \ref{fig:BPT_PlotTol1004-296MUSE} show both the BPT plots and the spatial distribution map from the MUSE data. Here it was possible to implement the third BPT plot, [OIII]$\lambda$5007 / H$\beta$ versus [OI]$\lambda$6300 / H$\alpha$. Two comments can be made when comparing the BTP plots between the data from MUSE and our data. First, as for the SIFS data, almost all points are located in the \HII\ zone, suggesting that what can be seen in the center of the galaxy (with SIFS FoV) is still true for the entire galaxy. Second, we can notice that the ratio is limited to $-1.5 < \log([\text{OIII}] \lambda 5007 / H\beta) < +1.0$. No points are greater than 1.0, when previously, some points had values up to +1.8. It is possible that those points have lower SNR in the outskirts of the SIFS FoV. The ratio $-3.0 < \log([\text{S II}] \lambda\lambda 6717, 6731 / H\alpha) < +0.1$, whereas the higher ratio for the SIFS data was -0.1. A few points are located into the LINER zones. The ratio $-4 < \log([\text{O I}] \lambda 6300/H\alpha) < -0.1$ confirms that the overwhelming majority of the ionization mechanism is coming from hot stars UV photons photoionization.

The distribution map in Figure \ref{fig:BPT_PlotTol1004-296MUSE} confirms what has been found with the SIFS data in the center. In the outskirts of the galaxy is further from the maximum starburst line.

\subsubsection{Tol 0957-278}

\begin{figure}
	\includegraphics[width=0.45\textwidth]{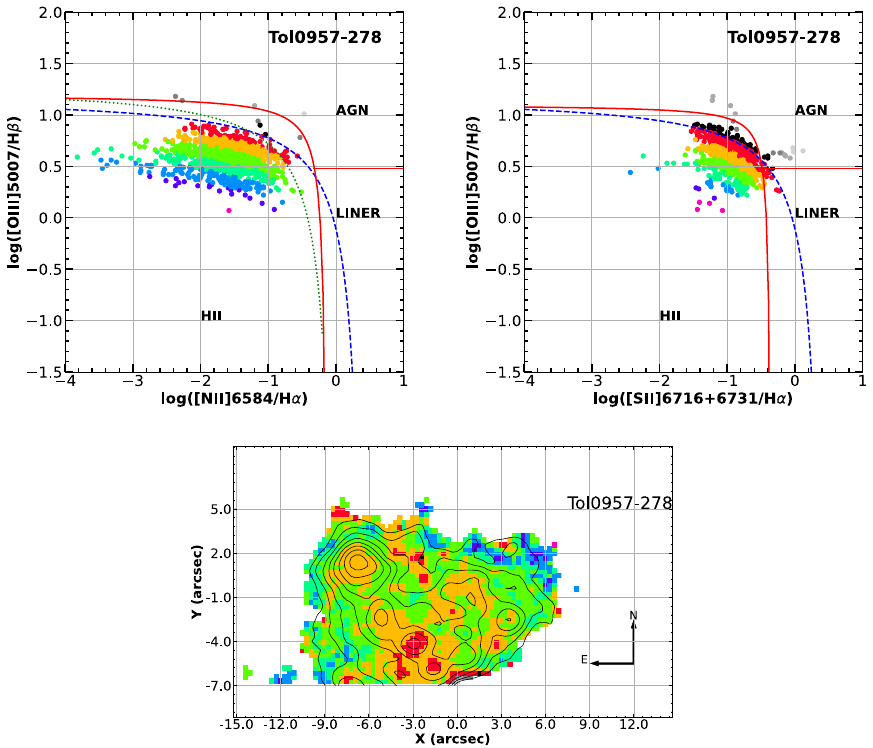}
    \caption{Same as Figure \ref{fig:BPT_PlotMapTol1004-296}, but for Tol 0957-278.} 
    \label{fig:BPT_PlotMapTol0957-278}
\end{figure}

On Figure \ref{fig:BPT_PlotMapTol0957-278} (Top), both plots show the expected result with pretty much all points located in the \HII\ zone. 
Line ratios range between  $-4.0 < \log([\text{NII}] \lambda 6584 / H\alpha) < -0.6$ and between $+0.2 < \log([\text{OIII}] \lambda 5007 / H\beta) < +0.8$.
The ratio $-2.0 < \log([\text{NII}] \lambda 6584 / H\alpha) < 0.0$ is also showing that most of the pixels are located in the \HII\ area, with only a few points above the \citet{Kewley2001} frontier.

The spatial distribution on Figure \ref{fig:BPT_PlotMapTol0957-278} also shows that pixels close to the maximum starburst line are concentrated on the bright emission knot 2 and 3.

\subsection{Advanced kinematical analysis}
The other interest of this study is the kinematical analysis aspect. In that regard, we observed both objects using the most resolving grating, M1500M3 available for SIFS. The M1500M3 grating gives a resolution of 9500@5500\AA\footnote{https://noirlab.edu/science/programs/ctio/instruments/sifs/characteristics}. The spectral sampling at H$\alpha$ is 0.2877\AA\ (13 km s$^{-1}$). Even with this grating, we also used the data observed with the R700M grating to produce velocity and velocity dispersion maps (not presented here) in order to check our findings. In this section, we will describe the velocity and velocity dispersion  maps for both galaxies. We also look at the kinematical diagnostic diagrams \citep{Bordalo2009}, applying statistical methods in order to determine possible systematic motions. We will also explore in more detail the line emission profiles in order to describe different degrees of symmetry. As a note, observations done using the M1500M3 grating do not match exactly the fields location showed in Figure \ref{fig:LTO_Tol1004_Tol0957}. Small differences can be noted.

\subsubsection{Kinematics of Tol 1004-296}
\begin{figure*}
	\includegraphics[width=\textwidth]{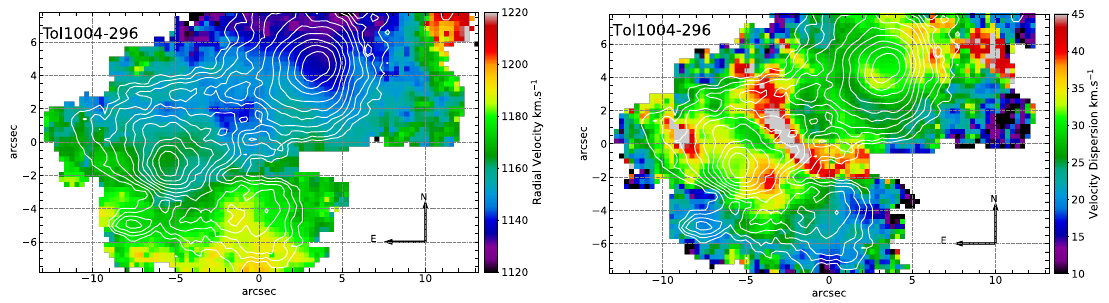}
    \caption{{\it Left} Radial velocity map. {\it Right} Velocity dispersion map. We superimposed H$\alpha$ isocontours on both maps.} 
    \label{fig:VrDispTol1004-296}
\end{figure*}

Figure \ref{fig:VrDispTol1004-296} shows the radial velocity and the velocity dispersion maps of Tol 1004-296. In both maps, we have superimposed the isocontours from the associated H$\alpha$ emission. The radial velocity ranges from 1137 $\pm$ 5 km s$^{-1}$ to 1205 $\pm$ 6 km s$^{-1}$, between the different knots, showing a velocity gradient from the South East to the North West. The velocity amplitude between Knot 1 and Knot 2 is 25 $\pm$ 4 km s$^{-1}$ (with Knot 1 being in the blue side). This velocity amplitude does not imply rotation. The velocity field does not show a disk pattern. It can be noticed that, within Knot 1, a small velocity gradient is also present ($\approx$ 15 $\pm$ 3 km s$^{-1}$). Knot 3 also shows a small velocity gradient ($\approx$ 10 $\pm$ 3 km s$^{-1}$). Also, Knot 5, in between the two larger Knots 1 and 2, shows a low radial velocity of 1147 $\pm$ 4 km s$^{-1}$. Higher radial velocities (from 1170 $\pm$ 4 km s$^{-1}$ to 1180  $\pm$ 4 km s$^{-1}$) can be seen in the South East and South, and the southern tip of Knot 4 shows the highest radial velocity of $\approx$1200 $\pm$ 6 km s$^{-1}$.

The velocity dispersion is also shown in Figure \ref{fig:VrDispTol1004-296}. Most of the galaxy shows a supersonic velocity dispersion, with values ranging from 17  $\pm$ 5 km s$^{-1}$ to 34 $\pm$ 2 km s$^{-1}$ between the different knots. The map shows that the velocity dispersion is not uniform across the galaxy. The centers of emission knots show slightly lower dispersion than the surroundings. This is true for Knot 1, Knot 2, and Knot 3. It is less obvious for Knot 5 and Knot 6. The other important thing to notice is the high velocity dispersion areas. Areas with a velocity dispersion higher than 40  $\pm$ 5 km s$^{-1}$ can be found in three locations. First, the area located at (-9.0\arcsec, +1.0\arcsec), the second one is located right between the two main knots with an elongated shape spanning $\approx 4.2 \arcsec$. And finally, an area located at (+10.0\arcsec, +5.0\arcsec) also shows velocity dispersion higher than 45 $\pm$ 7 km s$^{-1}$. But, in the last case, because the area is on the outskirts of the galaxy, the SNR is low and the profiles are not very well defined. The region in between the two Knots is the most unexpected with its linear shape. We looked at the data with the R700M grating in order to confirm this unusual shape. The shape, extension, and values are confirmed by those other data. We will come back in more detail on this region when studying the profiles.


Kinematical diagnostic diagrams, pioneered by \citet{MunozTunon1996} and \citet{Bordalo2009}, offer a robust method for analyzing gas motions in systems lacking disk rotation. These diagrams typically include plots of radial velocity versus emission intensity ($Vr$ vs. $Int$), velocity dispersion versus emission intensity ($\sigma$ vs. $Int$), and velocity dispersion versus radial velocity ($\sigma$ vs. $Vr$). Researchers have demonstrated the utility of these diagrams in studying turbulent motion in \HII\ galaxies \citep{Moiseev2012,Bordalo2009,Carvalho2018} and \HII\ regions \citep{Russeil2016}, enabling the identification of various kinematical features such as expanding shells and relative motion of regions within the object.

To extract information from these diagrams, particularly $\sigma$ vs. $Int$ and $\sigma$ vs. $Vr$, we employed a statistical method utilizing the R statistical package (R Development Core Team 2009). Specifically, we utilized the $Mclust$ function for model-based clustering, classification, and probability density estimation based on finite Gaussian mixture modeling \citep{Fraley2007}. The Bayesian Information Criterion (BIC) guided the selection of the number of mixture components and the best covariance parameterization. This approach facilitated the identification of independent populations within the diagrams, both in terms of their distribution within the diagrams and their geographical location on the map.

Rather than attempting to separate different components as done in a previous study \citep{Carvalho2018}, we opted to use the $Density$ option, which provides an estimate of the density (probability) for each point on the diagram using a finite mixture model from $Mclust$.

\begin{figure*}
	\includegraphics[width=\textwidth]{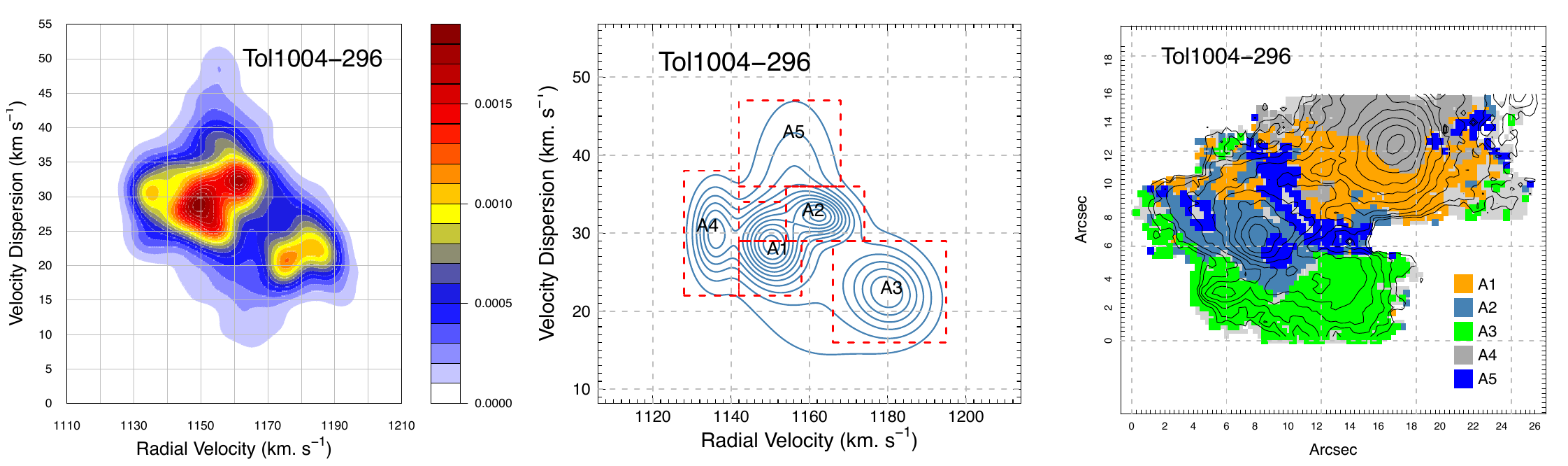}
    \caption{Kinematical Diagnostic Diagrams. Left: $\sigma$ vs Radial Velocity diagram using point density visualisation (the density color bar represents relative values normalized to one). Middle: $\sigma$ vs Intensity diagram after MClust analysis and with five regions identified. Right: Spatial localisation of the pixels inside the five regions.} 
    \label{fig:FigureRKinematicalAnalysis2Tol1004-296}
\end{figure*}

\begin{itemize}

\item $\sigma$ vs. Intensity:

Figure \ref{fig:FigureRKinematicalAnalysis1Tol1004-296} comprises three plots. On the left, we present the density map of the velocity dispersion versus intensity diagram, with the density color bar indicating relative values. This map highlights point concentrations with varying colors, with redder shades indicating higher concentration. The middle plot represents the results of a direct analysis with $Mclust$ in terms of probability density, identifying five zones of concentration ($A1, A2, A3, A4, A5$). While some similarities to previous diagrams exist, certain components, such as the inclined bands observed by \citet{Moiseev2012}, are not as prominent here. Notably, areas $A1$ and $A2$ correspond to regions of low turbulence in the interstellar medium (ISM), while $A4$ resembles the \HII\ region area in previous sketches. The plot on the right attempts to spatially locate the different points from the identified areas ($A1, A2, A3, A4, A5$) on an spatial location map. Dark grey shading represents $A4$, corresponding to high-emission areas such as Knot 1 and Knot 2. Areas $A1$ and $A2$ encompass peripheral pixels with lower signal-to-noise ratio (SNR), while $A3$ and $A5$ surround the two main knots. The $A2$ area identifies regions with lower velocity dispersion and medium intensity, including Knots 3, 6, and 7. To avoid overloading the map, points that are not in regions $A1$ to $A5$ are depicted in light grey. This includes regions with high (above 35 km s$^{-1}$) and low (below 18 km s$^{-1}$) velocity dispersion (low velocity dispersion are located on the outskirts of the map). \\

\begin{figure*}
	\includegraphics[width=\textwidth]{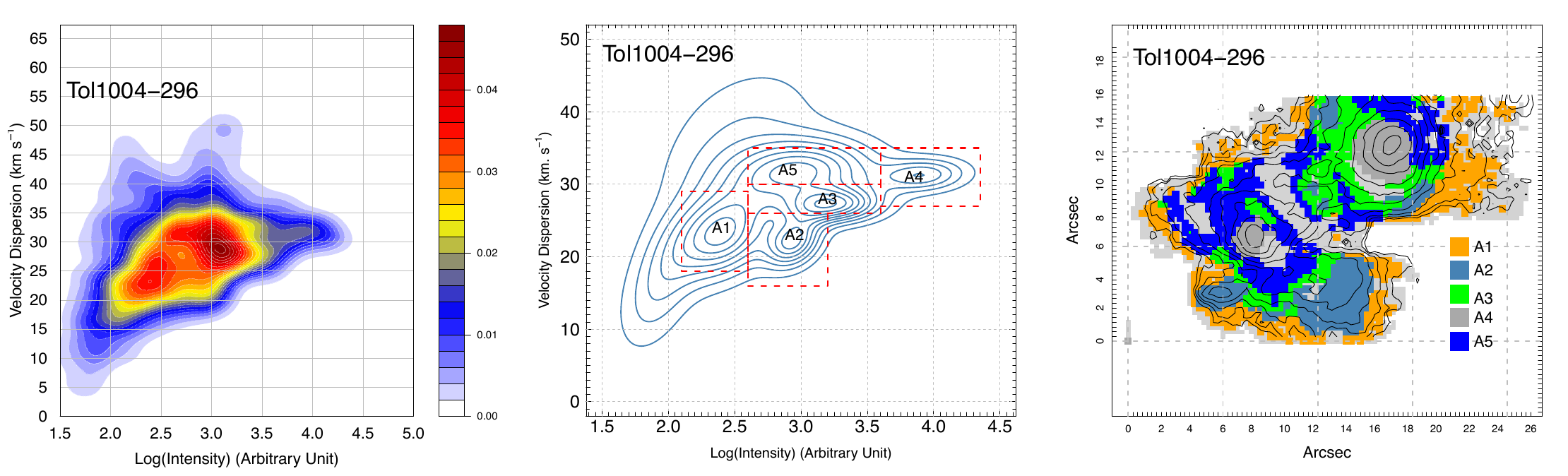}
    \caption{Kinematical Diagnostic Diagrams. Left: $\sigma$ vs Intensity diagram using point density visualisation (the density color bar represents relative values normalized to one). Middle: $\sigma$ vs Intensity diagram after MClust analysis and with five regions identified. Right: Spatial localisation of the pixels inside the five regions.} 
    \label{fig:FigureRKinematicalAnalysis1Tol1004-296}
\end{figure*} 

\item $\sigma~vs~Radial~Velocity$:

Figure \ref{fig:FigureRKinematicalAnalysis2Tol1004-296} represents the same plots as before, but for the $\sigma~vs~Radial~Velocity$ diagram. The left plot shows more clearly the different radial velocity regimes of the galaxy, with the main two concentrated areas representing Knot 1 and Knot 2. This can easily be seen on the velocity map. But the analysis of the probability density, presented in the middle shows that four areas, $A1, A2, A3, A4$ has been identified by $MClust$. We decide to include a fifth region, $A5$, representing the high velocity dispersion area. According to \citet{Bordalo2009}, the inclination of different regions in that diagram shows the systemic motion of the region one to the other. We can see that the different zones $A1, A2, A3$ have systemic motions in one direction. The case of $A4$ is difficult because the region seems to be vertical. The interpretation of this plot is helped by the next plot on the right. The analysis confirms the assessment we did from the velocity map saying that the Knot 1 shows a velocity gradient (regions $A1$ and $A4$ also showing the gradient within Knot 1). The analysis is also showing that the South- South East (Knot 1, Knot 6 and Knot 7) of the galaxy is moving away, but with a low velocity dispersion. The high velocity dispersion area coming from region $A5$ shows the straight pattern in the middle and area close to Knot 2, confirming the complexity of the dynamic of this Knot.

In conclusion, we can say that this analysis is confirming the analysis that can be done by looking to the different maps, by presenting, in a more clearer way, the different regimes. It also confirms that, in this kind of galaxies, dynamics is complex and is not dominated by the radial velocity nor the velocity dispersion, since the reconstruction of Figure \ref{fig:FigureRKinematicalAnalysis2Tol1004-296} is a mixture of both.

\end{itemize}

\subsubsection{Kinematics of Tol 0957-278}
\begin{figure*}
	\includegraphics[width=\textwidth]{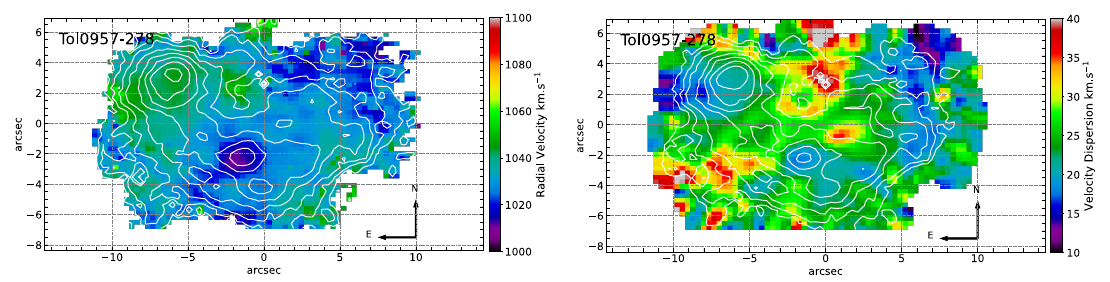}
    \caption{Radial velocity map. {\it Right} Velocity dispersion map. We superimposed H$\alpha$ isocontours on both maps.} 
    \label{fig:VrDispTol0957-278}
\end{figure*}

Figure \ref{fig:VrDispTol0957-278} shows the radial velocity and velocity dispersion maps of Tol 0957-278. Radial velocities range between 1016 $\pm$ 4 km s$^{-1}$ and 1041 $\pm$ 5 km s$^{-1}$, between the different regions. Unlike Tol 1004-296, the velocity field doesn't exhibit a consistent velocity gradient. However, it's notable that Knot 2 has a lower radial velocity (approximately 1016 km s$^{-1}$ $\pm$  5 km s$^{-1}$) compared to Knot 1, which has a much higher velocity (approximately 1041 $\pm$ 5 km s$^{-1}$). Additionally, a small area (located at -2\arcsec, +2.5\arcsec in Figure \ref{fig:VrDispTol0957-278}(Right)), not associated with an emission knot, shows the highest radial velocity (approximately 1050 $\pm$ 5 km s$^{-1}$).

The velocity dispersion map in Figure \ref{fig:VrDispTol0957-278} reveals non-uniformity, with dispersion values ranging from 20 $\pm$ 3 km s$^{-1}$ to 27 $\pm$ 3 km s$^{-1}$ between the different regions. Several emission knots (Knots 1, 2, 3, 4, and 5) exhibit lower velocity dispersion in their centers compared to their surroundings. There are some hot spots where the velocity dispersion is between 35 $\pm$ 3 km s$^{-1}$ and 40 $\pm$ 3 km s$^{-1}$. One is 6\arcsec\ West from Knot 1, another one is  3 \arcsec\ North West from  Knot 2 and finally one 6 \arcsec\ South East from Knot 1.

In the following sections, we present the same statistical analysis as for Tol 1004-296.
 
\begin{figure*}
	\includegraphics[width=\textwidth]{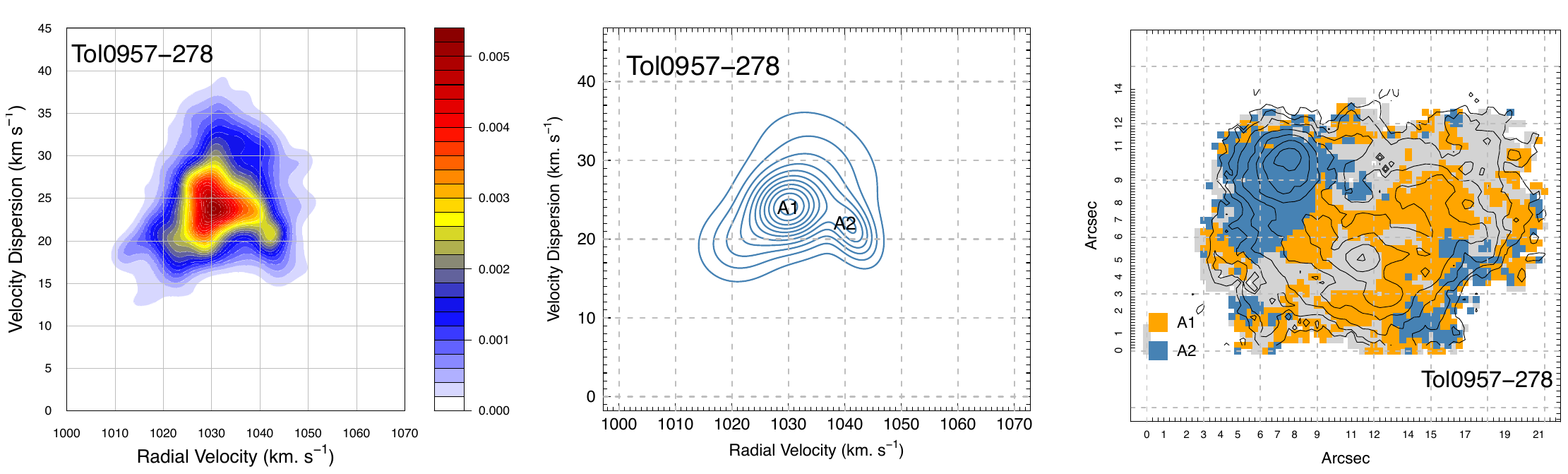}
    \caption{$\sigma~vs~Vr$ kinematical diagnostic diagrams for Tol 0957-278 (same as Tol 1004-296)} 
    \label{fig:FigureRKinematicalAnalysis2Tol0957-278}
\end{figure*}

\begin{itemize}

\item $\sigma~vs~Intensity$:

Figure \ref{fig:FigureRKinematicalAnalysis1Tol0957-278} presents the same three plots as before. On the left is the $\sigma~vs~Intensity$ diagram, where the concentration of points is color-coded. The velocity dispersion ranges between approximately 14 km s$^{-1}$ and 40 km s$^{-1}$. The overall shape of this plot closely resembles the sketch presented by \citet{Moiseev2012}, with high-intensity regions exhibiting nearly constant sigma (here approximately 22 km s$^{-1}$). The area referred to as low turbulence ISM can also be observed, with sigma ranging from approximately 14 km s$^{-1}$ to 40 km s$^{-1}$ and intensity between 1.7 to 2.5. Finally, a possible inclined band (mentioned as A3 in Figure \ref{fig:FigureRKinematicalAnalysis1Tol0957-278} middle) can be discerned, but we will revisit this later.

The middle plot represents the probability density from the $Mclust$ analysis, identifying three concentration areas: $A1$, $A2$, and $A3$. We add two more regions, $A4$ representing the high-intensity emission regions, and $A5$ representing high velocity dispersion and low intensity. The plot on the right depicts the locations of the points from the different areas $A1$ to $A5$. The $A4$ area can be identified with the main emission knots: Knot 1, Knot 2, and Knot 3. $A1$ represents points on the outskirts of the galaxy. The $A5$ area, representing high velocity dispersion, does not correspond to a particular localization (except perhaps around Knot 8). The location of $A3$ pixels suggests that it is unlikely to represent an expanding shell as suspected before.\\

\begin{figure*}
	\includegraphics[width=\textwidth]{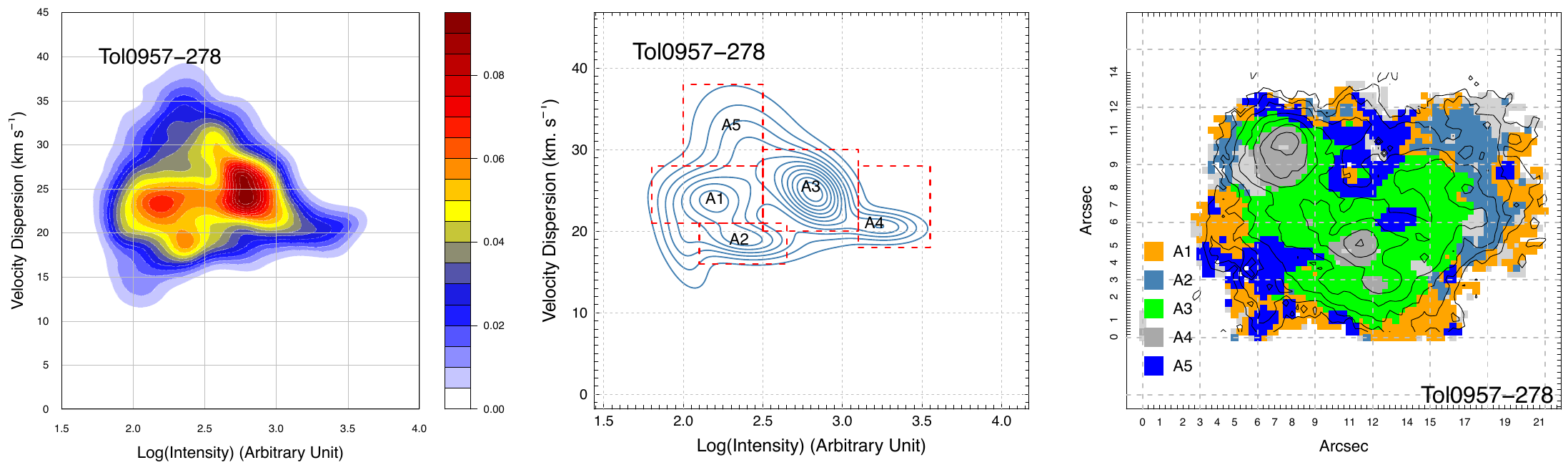}
    \caption{$\sigma~vs~Int$ kinematical diagnostic diagrams for Tol 0957-278 (same as Tol 1004-296)} 
    \label{fig:FigureRKinematicalAnalysis1Tol0957-278}
\end{figure*} 

\item $\sigma~vs~Radial~Velocity$:

Figure \ref{fig:FigureRKinematicalAnalysis2Tol0957-278} presents the same three plots as before. On the left is the $\sigma~vs~Radial~Velocity$ diagram, with the concentration of points color-coded. The $MClust$ result is shown in the middle plot, and no clearly independent regions are present. We only present two areas, $A1$ and $A2$, from the probability density analysis. The location of pixels from $A1$ and $A2$ is shown in the right plot. $A1$ represents Knot 1 and the surrounding area, while the $A2$ area encompasses the region around Knot 2. This analysis is not very informative, as the $\sigma~vs~Radial~Velocity$ diagram only exhibits two populations, and the analysis does not offer much insight in this case.

\end{itemize}

\subsection{Profile Analysis}
In this section, we examine the shapes of various H$\alpha$ line profiles across both objects in greater detail. We present 11 profiles from different areas of each galaxy, with each profile representing the sum of a 3x3 pixel region (approximately $0.9\arcsec^2$). The objective of this section is to understand the supersonic nature of these objects, particularly whether the very high velocity dispersion (above 30 km s$^{-1}$) is due to the presence of multiple components or is intrinsically large.

\subsubsection{Tol 1004-296}
\begin{figure*}
	\includegraphics[width=\textwidth]{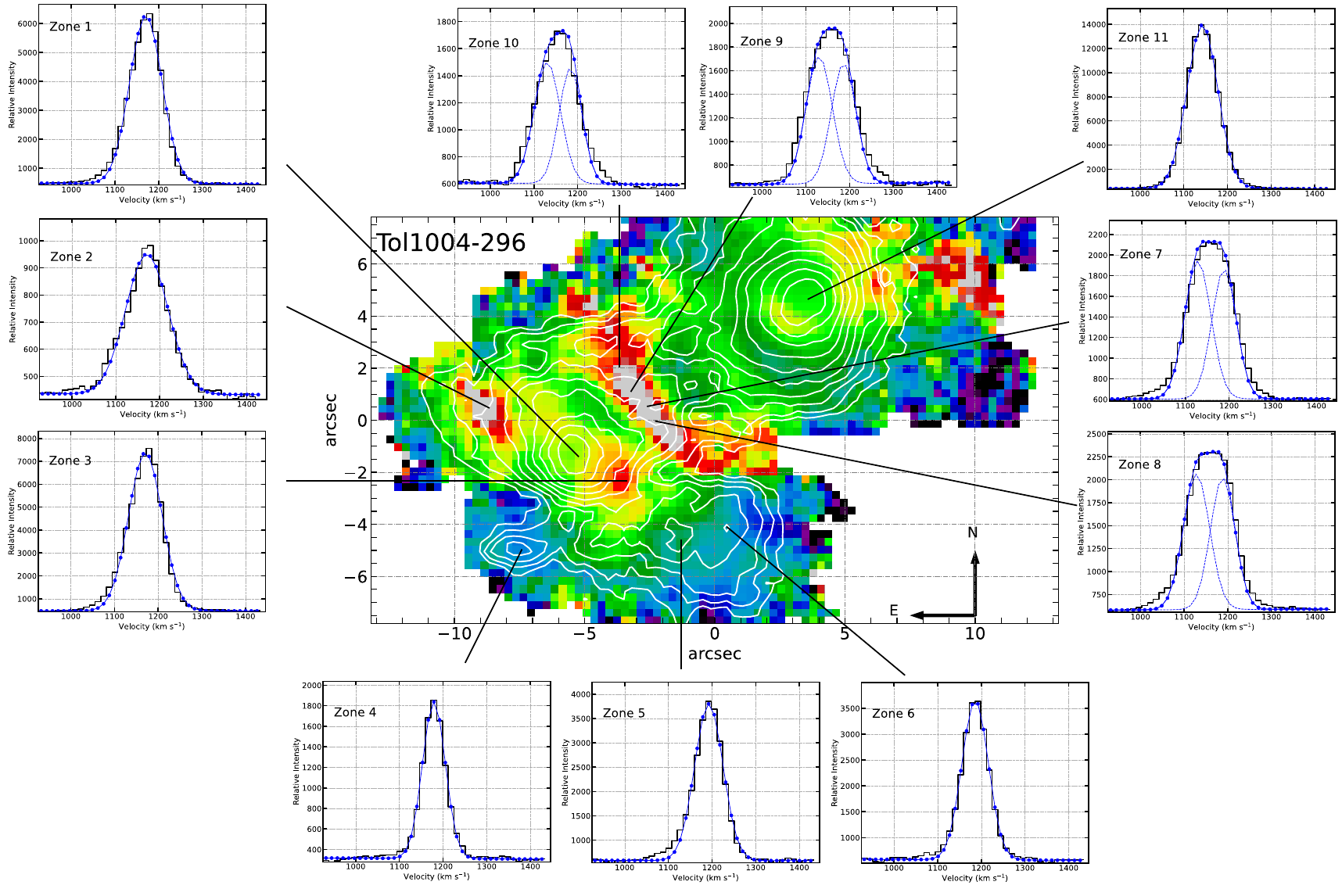}
    \caption{Presentation of H$\alpha$ profiles in different part of Tol 1004-296. Each Zone (1 to 11) represents the integrated profiles in a 3x3 pixel box (representing $\approx 0.9\arcsec^2$) and represents the intensity versus the radial velocity. Except for Zone 7, Zone 8, Zone 9 and Zone 10, where it was possible to fit two gaussian, the rest of the profiles could be fit with only one gaussian. H$\alpha$ emission isocontours are superimposed over the velocity dispersion of the galaxy.} 
    \label{fig:Tol1004-296Profiles}
\end{figure*}

Figure \ref{fig:Tol1004-296Profiles} displays the H$\alpha$ emission profiles in different areas of Tol 1004-296. The selected areas include regions where the emission is at its maximum (Zones 1, 4, 5, 6, and 11), where the velocity dispersion is highest (Zones 2, 7, 8, 9, and 10), and an area where the velocity dispersion is high but the emission is not at its maximum (Zone 3).

Upon examining the profiles in Zones 1, 4, 5, 6, and 11, it is evident that a single Gaussian adequately fits the emission profile. However, the left wing of Zone 5 is not well-fitted by the Gaussian, suggesting the potential presence of a second component with very low amplitude and velocity. Zones 2 and 3 appear to exhibit the presence of a second component on the low-velocity side.

Zones 7 to 10 encompass the region between the two bright knots (Knot 1 and Knot 2) with a linear shape and higher velocity dispersion. Although these profiles are fairly symmetrical, it is apparent that such broad profiles can be separated into two components. In Figure \ref{fig:Tol1004-296Profiles}, we attempt such a decomposition for Zones 7 to 10. The results indicate that two components with approximately equal amplitude and sigma can replicate the original profile. However, it is also noticeable that the wings, especially in the low-velocity range, are not well-fitted. Introducing a very low amplitude and low-velocity third component may be necessary to accurately reproduce these profiles. However, we refrain from this analysis as a profile can be adequately fitted with numerous components; the crucial aspect is the ability to track these components from profile to profile.

Several observations can be made from this analysis. Firstly, the centers of the different knots can be fitted with only one Gaussian. Secondly, the linear region where the velocity dispersion is highest clearly displays two components (of more or less identical intensities and sigmas) and is indicative of ongoing interactions.

\subsubsection{Tol 0957-278}
\begin{figure*}
	\includegraphics[width=\textwidth]{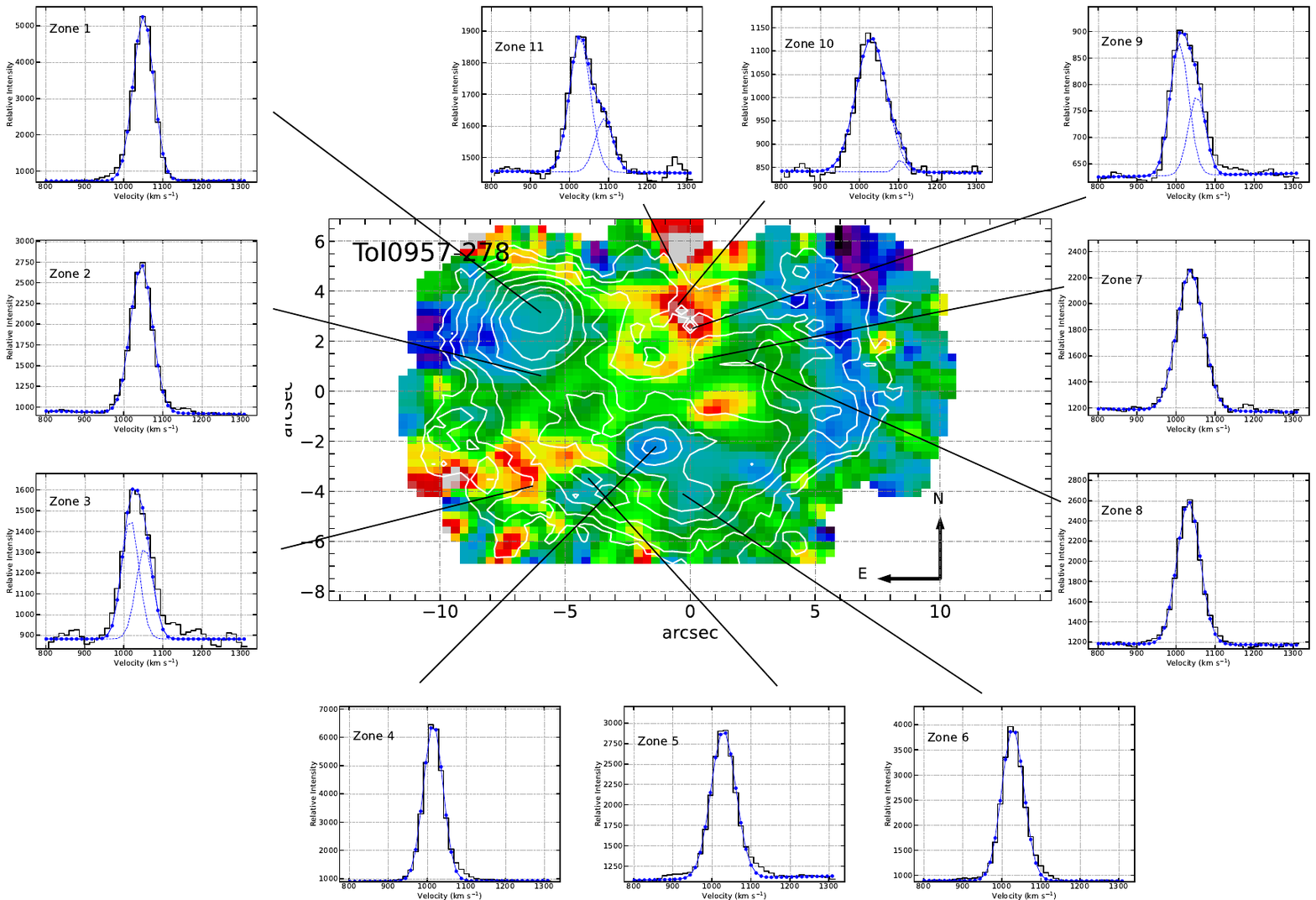}
    \caption{Same as Tol 1004-296. Here Zone 3, Zone 9, Zone 10 and Zone 11 profiles, have been fitted with two gaussian. Contrary to Tol 1004-296, there is a high and low amplitude gaussian.} 
    \label{fig:Tol0957-278Profiles}
\end{figure*}

Figure \ref{fig:Tol0957-278Profiles} presents the H$\alpha$ emission profiles from different areas of Tol 0957-278. Zones 1, 4, 5, and 6 represent integrated profiles from emission knots, while Zones 2, 7, and 8 display profiles from the surroundings of knots. Zones 3, 9, 10, and 11 correspond to regions with high velocity dispersion.

Profiles from Zones 1, 4, 5, and 6 can be adequately fitted with only one component, as depicted in Figure \ref{fig:Tol0957-278Profiles}, although Zone 4, Zone 5, and Zone 6 profiles exhibit a low-intensity component on the right wing. Similarly, profiles in Zones 2, 7, and 8 have been fitted with one Gaussian, albeit with a slightly larger velocity dispersion. However, profiles from Zones 3, 9, 10, and 11 display asymmetric shapes and have been fitted with two components, with one having a higher intensity than the other (but with sigmas remaining more or less the same).

Similar to Tol 1004-296, a low-intensity third component is observed in some profiles. Large profiles from Zones 9 to 11 can be decomposed into two Gaussians, while profiles from low emission areas are primarily fitted with one component.

Additionally, we can note a more pronounced asymmetry with profiles in Zones 9 to 11.

\subsection{Line ratios and Velocity Dispersion}
\begin{figure*}
	\includegraphics[width=\textwidth]{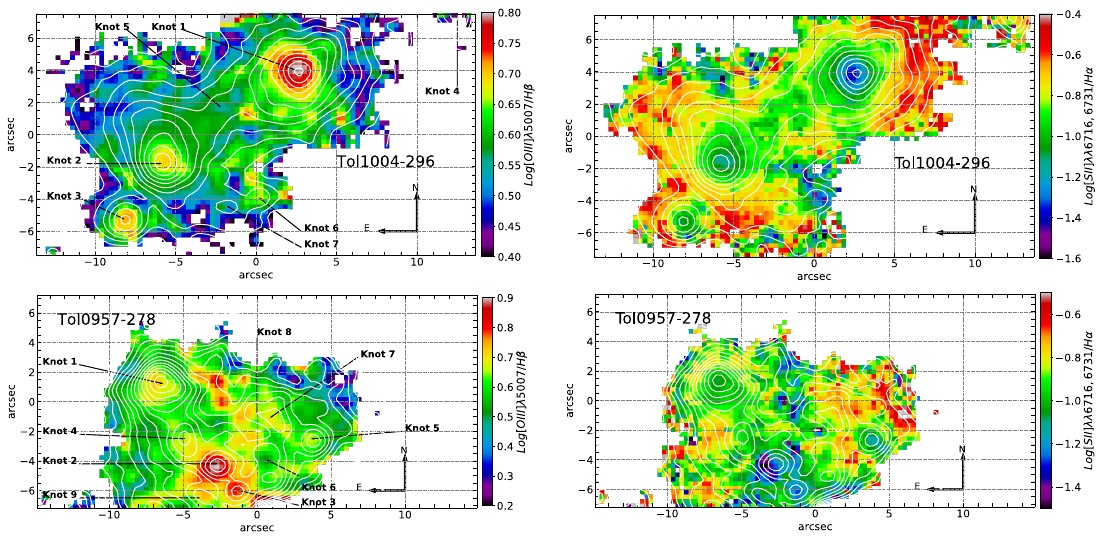}
    \caption{First row Tol 1004-296- Left: $log([OIII]5007/H\beta)$ line ratio. Right: $log([SII]/H\alpha)$ line ratio. Isocontours represent $H\alpha$ emission.
    Second row Tol 0957-278- Left: $log([OIII]5007/H\beta)$ line ratio. Right: $log([SII]/H\alpha)$ line ratio. Isocontours represent $H\alpha$ emission.}
    \label{fig:LinesRatiosSigma}
\end{figure*}

In this section, we begin to investigate a possible correlation between different line ratios and velocity dispersion, following previous works by \cite{Oparin2018} and \cite{DAgostino2019}. In Figure \ref{fig:LinesRatiosSigma}, we present line ratio maps of $log([OIII]5007/H\beta)$ and $log([SII]/H\alpha)$ for both objects, which are to be compared with the velocity dispersion in Figures \ref{fig:VrDispTol1004-296} and \ref{fig:VrDispTol0957-278}.

$[OIII]5007/H\beta$ is sensitive to the radiation field and metallicity, with a high number of high-energy photons producing high ionization, while $[SII]/H\alpha$ is sensitive to shocks. In the $log([OIII]5007/H\beta)$ maps of Tol 1004-296, clear peaks of the ratio are observed in the main emission knots (1, 2, 3, 4, 6, 7), with the ratio diminishing around them. Conversely, the $[SII]/H\alpha$ ratio shows the exact opposite pattern in the same knots, as expected in regions ionized by massive hot stars. Areas west and northwest from Knot 1, as well as an area south of Knot 1, show $log([SII]/H\alpha)$ > -0.3, indicating another ionization mechanism, possibly shocks. These maps should also be compared with the velocity dispersion map. Lower velocity dispersions correspond to high values of the $log([OIII]5007/H\beta)$ ratio and low values of the $log([SII]/H\alpha)$ ratio (located in high emission knots). This correspondence can also be extended to the metallicity map in Figure \ref{fig:MetallicityTol1004-296_Tol0957-278}. As mentioned before, metallicity is lower in the star-forming sites. In the area where Tol 1004-296 exhibits a peculiar linear pattern in the velocity dispersion map, the $log([OIII]5007/H\beta)$ ratio, though not at its maximum as in Knots 1 and 2, is fairly high, with a value between  0.58 $\pm$ 0.02 and 0.62 $\pm$ 0.02. The $log([SII]/H\alpha)$ ratio in the same area ranges between -0.78 $\pm$ 0.07 and -0.86 $\pm$ 0.05, higher compared to the very center of the knots. This linear feature is likely related to the collective mechanical energy from the stellar winds and the supernovae explosions sweeping up the ISM in opposite direction from the center of the two main star-forming knots. 
The high $log([OIII]5007/H\beta)$ and low $log([SII]/H\alpha)$ is characteristic of star-forming regions of Tol 0957-278. For the Knot 2, the ionization level, as given by $log([OIII]5007/H\beta)$, is high compared to Knot 1. Combining this information with the low oxygen abundance observed in Knot 2, one may conclude that there must have been a recent inflow of metal-poor gas to this region feeding the ongoing star 
formation. This is also in agreement with the low velocity dispersion in the star forming knots.

\section{Conclusions}

In this work, we present IFU observations of two \HII\ galaxies, Tol 1004-296 and Tol 0957-278. The lower spectral resolution observations allow us to produce several main emission line monochromatic maps, along with electron density, reddening, line ratios, and metallicity maps. We also had the possibility to explore the ionization mechanism using BPT diagrams.  Higher spectral resolution data enable the generation of precise radial velocity and velocity dispersion maps.

In this work, we delved deeper into the analysis of the kinematics by using kinematical diagnostic diagrams to show if systemic motions, expanding shells were present in these two objects and how this can help to understand the characteristics of the ISM.

Both galaxies show the common features we expect for \HII\ galaxies. They present several bright emissions peaks in the different lines. We count seven and eight knots, respectively, for Tol 1004-296 and Tol 0957-278. Extinction, due to dust lane, can be severe, as in Knot 1 in Tol 0957-278. The metallicity does not exceed 8.5, and the ionization mechanism is dominated by photoionization from OB stars.

The kinematical perspective confirms the image we usually have of such objects: no disk rotation, but a slight velocity gradient is present between the two main knots in the Tol 1004-296 case. However, the velocity dispersion shows its importance when we applied a more in-depth analysis using statistical tools to analyze the diagnostics diagrams. We didn't find expanding shells, but systemic motions are present with signatures of turbulence. The profile analysis of the H$\alpha$ line profile shows that emission is made of one component where the emission peaked and of two components in some specific areas, where the velocity dispersion is higher, showing possible turbulent motions there.

We can categorize the results of this study into two main findings.

Firstly, we confirm a correlation observed in previous studies (see Table \ref{tab:summarytab}), wherein a peak of emission corresponds to low velocity dispersion. This result can be extended to $[OIII]5007/H\beta$, $[SII]/H\alpha$ line ratios maps, and metallicity maps. We found that peak emission regions (knots) correspond to maximum $[OIII]5007/H\beta$ ratio, minimum $[SII]/H\alpha$ ratio. Additionally empirical metallicity maps revealed lower metallicity in regions with low velocity dispersion. In Tol 1004-296, we verified this using three different calibrators, albeit with slightly different values. Similarly, in Tol 0957-278, the metallicity map reflects lower values where the velocity dispersion is low. It has to be noted that these correspondences are present in almost all previous studies (see Table \ref{tab:summarytab}). Authors of these studies mentioned the correspondence between peak emission with peak $[OIII]5007/H\beta$ ratio, low $[SII]/H\alpha$, and sometimes with low metallicity, but, to our knowledge, none of these studies have connected this with a low velocity dispersion. We summarize these findings in Table \ref{tab:summarytab}, which lists 16 objects (including this study) exhibiting such characteristics.

Secondly, we note a significant result regarding the velocity dispersion map of Tol 1004-296. Figure \ref{fig:VrDispTol1004-296} clearly depicts a linear region with high velocity dispersion ($\approx$ 50 $\pm$ 5  km s$^{-1}$) situated between the brighter emission knots, a feature also observed in data obtained from the R700M grating. The decomposition of these profiles in Figure \ref{fig:Tol1004-296Profiles}, albeit crude, reveals two distinct Gaussians. Notably, we did not encounter a similar feature in the literature's velocity dispersion maps. While we lack a definitive interpretation or explanation for this phenomenon, one possible explanation is the collision of material pushed from both Knot 1 and Knot 2, leading to the formation of a region with high velocity dispersion.

Results regarding the different line ratios in the knots are consistent with the image of young massive star formation sites, with high $[OIII]5007/H\beta$ ratio, low $[SII]/H\alpha$ ratio and low metallicity. The fact that the velocity dispersion is also lower in these areas compared to the surroundings, tend to show that turbulent motions are propagated around the area where the emission is at its maximum. This seems to be corroborated by the $[SII]/H\alpha$ ratio, sensitive to shocks.

\begin{table}
	\caption[]{Summary of key galaxies characteristics of a sample of BCGs: (1) This work; (2) \cite{Cairos2015}; (3) \cite{Cairos2020}; (4) \cite{Cairos2022}; (5) \cite{Egorov2021}; (6) \cite{Castillo-Morales2011}; (7)\cite{GarciaLorenzo2008}\label{tab:summarytab}}
	     \tiny
		\begin{tabular}{lccccc}
		\hline\hline
Galaxy & High                        & High                &  Low                   & Velocity     & Rotating    \\.    
            & emission.                 & emission vs    &  Metallicity          &                  &                  \\
            & vs Low $\sigma $    &    Low             &  vs Low $\sigma $                   & Gradient   &    Disk        \\
             &                                &   Metallicity     &                            &                  &                  \\
                       \hline\hline
	   Tol 1004-296$^{(1)}$       & Yes & Yes & Yes & Yes & No\\ 
	   Tol 0957-278$^{(1)}$       & Yes & Yes & Yes & No & No\\
	   Haro 11$^{(2)}$            & Yes & Yes & Yes & No & No\\
	   Haro 14$^{(2,4)}$	           & Yes & Yes & Yes & No & No\\
	   Tol 0127-397$^{(2)}$	   & Yes & Yes & Yes & Yes & No\\
    	   Tol 1924-416$^{(2)}$	   & Yes & Yes & Yes & No & No\\
	   Tol 1937-423$^{(2)}$	   & Yes & Yes & Yes & Yes & Yes\\
	   Mrk 900$^{(2,3)}$            & Yes & Yes & Yes & Yes & No\\
	   Mrk 1131$^{(2)}$           & Yes & Yes & Yes & Yes & Yes\\
	   DDO 53$^{(5)}$             & Yes & Yes & Yes & No & No\\
	   NGC 7673$^{(6)}$          & --$^{(a)}$ & Yes & Yes &  -- & -- \\
	   Mrk 370$^{(7)}$            & Yes & Yes & Yes & Yes & No\\
            Mrk 35$^{(7)}$             & Yes & Yes & Yes & Yes & No\\
            Mrk 297$^{(7)}$            & Yes & Yes & Yes & No & No\\
            Mrk 314$^{(7)}$            & Yes & Yes & Yes & No & No\\
            III Zw 102$^{(7)}$         & Yes & Yes & Yes & Yes & No\\ 
           \hline\hline \\
           $^{(a)}$ no dispersion map available
 		\end{tabular}
\end{table}

\section*{Acknowledgements}
Based on observations obtained at the Southern Astrophysical Research (SOAR) telescope, which is a joint project of the Minist\'erio da Ciência, Tecnologia, e Inova\c c\~ao (MCTI) da Rep\'ublica Federativa do Brasil, the U.S. National Optical Astronomy Observatory (NOAO), the University of North Carolina at Chapel Hill (UNC), and Michigan State University (MSU). We thank LNA staff for his support during the runs and for dara reduction.
We acknowledge the usage of the Nasa Extragalactic Database (http://ned.ipac.caltech.edu/) and R free software. The authors also thank J.C. Lambert (LAM) for the help provided in the creation of certain figures. HP thanks the Laboratoire d'Astrophysique de Marseille (LAM) and Aix Marseille Universit\'e (AMU) for the financial support during his staying at LAM in December 2023. VGA thanks Funda\c c\~ao de Amparo e Pesquisa do Estado da Bahia (FAPESB) for the financial under fellowship BOL0345/2021.
Authors would like to thank the referee for their helpful remarks

\section*{Data Availability}
The data underlying this article will available in the NOIRLab Astro Data Archive (https://astroarchive.noirlab.edu/) or upon request to the authors. 

 



\begin{thebibliography}{99}

\bibitem[\protect\citeauthoryear{Alloin et al$.$}{1979}]{Alloin1979} Alloin, D., Collin-Souffrin, S., Joly, M.,  Vigroux, L. 1979, \aa, 78, 200

\bibitem[\protect\citeauthoryear{Ambartsumian et al$.$}{1968}]{Ambartsumian1968} Ambartsumian, V.A. 1968, L'Astronomie, Vol. 82, p.101

\bibitem[\protect\citeauthoryear{Amram}{1991}]{Amram1991} Amram, P. 1991, Thesis, Universit\'e de Provence: Cinématique du gaz ionisé dans les galaxies spirales. Effets de l'environnement dans les galaxies binaires et dans les amas

\bibitem[\protect\citeauthoryear{Bacon et al$.$}{2010}]{Bacon2010} Bacon, R., Accardo, M., Adjali, L., Anwand, H., Bauer, S., Biswas, I., Blaizot, J. et al. 2010, Proceedings of the SPIE, Volume 7735, id. 773508

\bibitem[\protect\citeauthoryear{Baldwin et al$.$}{1981}]{Baldwin1981}  Baldwin, J. A., Phillips, M. M., Terlevich, R. 1981, \pasp,  93, 5

\bibitem[\protect\citeauthoryear{Bergvall \& Olofsson}{1986}]{Bergvall1986} Bergvall, N., Olofsson, K., 1986., \aaps, 64, 469

\bibitem[\protect\citeauthoryear{Bordalo et al.}{2009}]{Bordalo2009} Bordalo, V., Plana, H., Telles,  E., 2009, \apj, 696, 1668

\bibitem[\protect\citeauthoryear{Bordalo \& Telles}{2011}]{Bordalo2011} Bordalo, V., Telles, E., 2011 \apj, 735, 52

\bibitem[\protect\citeauthoryear{Brennan et al$.$}{2017}]{Brennan2017} Brennan, R., Pandya, V., Somerville, R.S., Barro, G., Bluck, A.F.L., Taylor, E.N., Wuyts, S., Bell, E.F., Dekel, A., Faber, S. et al. 2017, \mnras, 465,  619

\bibitem[\protect\citeauthoryear{Buckalew et al$.$}{2005}]{Buckalew2005}  Buckalew, B.A., Kobulnicky, H.A., Dufour, R,J., 2005, \apjs, 157, 30

\bibitem[\protect\citeauthoryear{Calzetti et al$.$}{1994}]{Calzetti1994} Calzetti, D., Kinney, A.L., Storchi-Bergmann, T., 1994, \apj, 429, 582

\bibitem[\protect\citeauthoryear{Calzetti et al$.$}{2000}]{Calzetti2000} Calzetti, D., Armus, L., Bohlin, R.C., Kinney, A.L., Koornneef, J., Storchi-Bergmann, T.,  2000, \apj, 533, 682

\bibitem[\protect\citeauthoryear{Campbell et al$.$}{1988}]{Campbell1988} Campbell, A. 1988, \apj, 335, 644

\bibitem[\protect\citeauthoryear{Carvalho \& Plana}{2018}]{Carvalho2018}  Carvalho, M.S., Plana, H., 2018, \mnras, 481, 122 

\bibitem[\protect\citeauthoryear{Castillo-Morales et al$.$}{2011}]{Castillo-Morales2011} Castillo-Morales, A., Gallego, J., Pérez-Gallego, J., Guzmán, R., Muñoz-Mateos, J.C., Zamorano, J., Sánchez, S.F., 2011, \mnras, 411, 1819

\bibitem[\protect\citeauthoryear{Cairós et al$.$}{2003}]{Cairos2003} Cairós, L.M., García-Lorenzo, B., Caon, N., Vílchez, J.M., Papaderos, P., Noeske, K., 2003, \apss,  284, 611

\bibitem[\protect\citeauthoryear{Cairós et al$.$}{2015}]{Cairos2015} Cairós, L.M., Caon, N., Weilbacher, P.M., 2015, \aap, 577, A21

\bibitem[\protect\citeauthoryear{Cairós et al$.$}{2017a}]{Cairos2017a} Cairós, L.M., González-Pérez, J.N., 2017, \aap, 600, A125

\bibitem[\protect\citeauthoryear{Cairós et al$.$}{2017b}]{Cairos2017b} Cairós, L.M., González-Pérez, J.N., 2017, \aap, 608, A119

\bibitem[\protect\citeauthoryear{Cairós et al$.$}{2020}]{Cairos2020}  Cairós, L.M., González-Pérez, J.N., 2020, \aap, 634, A95    

\bibitem[\protect\citeauthoryear{Cairós et al$.$}{2022}]{Cairos2022} Cairós, L.M., Caon, N., Weilbacher, P.M., Manso Sainz, R., 2022, \aap, 164, A144

\bibitem[\protect\citeauthoryear{Chanda et al$.$}{2004}]{Chandar2004} Chandar, R., Leitherer, C., Tremonti, C.A., 2004, \apj, 604, 153 

\bibitem[\protect\citeauthoryear{Chhatkuli et al$.$}{2023}]{Chhatkuli2023} Chhatkuli, Daya Nidhi, Paudel, Sanjaya, Bachchan, Rajesh Kumar, Aryal, Binil, Yoo, Jaewon, 2023, \mnras, 520, 4953

\bibitem[\protect\citeauthoryear{Conti \& Morris}{1990}]{Conti1990} Conti P.S. and Morris P.W. 1990 \aj 99, 898

\bibitem[\protect\citeauthoryear{Cresci et al$.$}{2010}]{Cresci2010} Cresci, G., Vanzi, L., Sauvage, M., Santangelo, G., van der Werf, P., 2010, \aap, 520, A82

\bibitem[\protect\citeauthoryear{Cuisinier et al$.$}{2006}]{Cuisinier2006} Cuisinier, F., Westera, P., Telles, E., Buser, R., 2006, \aap, 455, 825

\bibitem[\protect\citeauthoryear{D'Agostino et al$.$}{2019}]{DAgostino2019}  D'Agostino, J.J., Kewley, L.J., Groves, B.A.,  Medling, A., Dopita, M.A., Thomas, A.D., 2019, \mnras, 485, L38

\bibitem[\protect\citeauthoryear{Denicoló et al$.$}{2002}]{Denicolo2002}  Denicoló, G., Terlevich, R., Terlevich, E., 2002, \mnras, 330, 69

\bibitem[\protect\citeauthoryear{Dinerstein et al$.$}{1990}]{Dinerstein1990} Dinerstein, H.L., in The interstellar medium in galaxies; Proceedings of the 2nd Teton Conference, Dordrecht, Netherlands, Kluwer Academic Publishers, 1990, p. 257-285.

\bibitem[\protect\citeauthoryear{Domínguez et al$.$}{2013}]{Dominguez2013} Domínguez, A., Siana, B.; Henry, A.L., Scarlata, C., Bedregal, A.G., Malkan, M., Atek, H., Ross, N.R. et al. 2013, \apj, 763, 145 

\bibitem[\protect\citeauthoryear{Dopita et al$.$}{2003}]{Dopita2003} Dopita, M.A., Groves, B.A., Sutherland, R.S., Kewley, L.J., 2003, \apj, 583, 727

\bibitem[\protect\citeauthoryear{Doublier et al$.$}{1999}]{Doublier1999} Doublier, V., Caulet, A., Comte, G., 1999, \aaps, 138, 213  

\bibitem[\protect\citeauthoryear{Egorov et al$.$}{2021}]{Egorov2021} Egorov, O.V., Lozinskaya, T.A., Vasiliev, K.I., Yarovova, A.D., Gerasimov, I.S., Kreckel, K., Moiseev, A,V., 2021, \mnras, 508, 2650

\bibitem[\protect\citeauthoryear{Eisenhauer et al$.$}{2003}]{Eisenhauer2003} Eisenhauer, F., Abuter, R., Bickert, K., Biancat-Marchet, F., Bonnet, H., Brynnel, J., Conzelmann, R.D., Delabre, B. et al. 2003, SPIE, 4841, 1548E

\bibitem[\protect\citeauthoryear{Engelbrachtet al.}{2008}]{Engelbracht2008} Engelbracht, C.W., Rieke, G.H., Gordon, K.D., Smith, J.D.T., Werner, M.W., Moustakas, J., Willmer, C.N.A., Vanzi, L., 2008, \apj, 678, 804

\bibitem[\protect\citeauthoryear{Fraga }{2018}]{Fraga2018} Fraga, L., Overall status and results from the commissioning and early science of SIFS. [s.n.], 2018, www.soartelescope.org/soar/ sites/default/files/SIFSFraga GMT2018.pdf

\bibitem[\protect\citeauthoryear{Fraley \& Raftery}{2007}]{Fraley2007} Fraley, C., Raftery, A.E., 2007, Journal of Statistical Software, v. 18, issue 6 

\bibitem[\protect\citeauthoryear{García-Lorenzo et al$.$}{2008}]{GarciaLorenzo2008} García-Lorenzo, B.; Cairós, L.M., Caon, N.; Monreal-Ibero, A.; Kehrig, C., 2008, \apj, 677, 201

\bibitem[\protect\citeauthoryear{Gil de Paz et al$.$}{2003}]{GildePaz2003} Gil de Paz, A., Madore, B.F., Pevunova, O., 2003, \apjs, 147, 29

\bibitem[\protect\citeauthoryear{Gil de Paz et al$.$}{2005}]{GildePaz2005} Gil de Paz, A., Madore, B.F., 2005,  \apjs, 156, 345

\bibitem[\protect\citeauthoryear{Greis et al$.$}{2016}]{Greis2016} Greis, S.M.L., Stanway, E.R., Davies, L.J.M.,Levan, A.J., 2016, \mnras, 459, 2591 

\bibitem[\protect\citeauthoryear{Hamuy et al$.$}{1993}]{Hamuy1992} Hamuy, M.; Walker, A.R., Suntzeff, N.B., Gigoux, P., Heathcote, S.R., Phillips, M.M., 1992, \pasp, 104, 533

\bibitem[\protect\citeauthoryear{Haro et al$.$}{1956}]{Haro1956} Haro, G., 1956, \aj, 61, 178

\bibitem[\protect\citeauthoryear{Hunter \& Elmegreen}{2004}]{Hunter2004} Hunter, D.A.,Elmegreen, B.G., 2004, \aj, 128, 2170

\bibitem[\protect\citeauthoryear{Kauffman et al$.$}{2003}]{Kauffman2003} Kauffmann, G., Heckman, T.M., Tremonti, C., Brinchmann, J., Charlot, S., White, S.D.M., Ridgway, S.E., Brinkmann, J., Fukugita, M., Hall, P.B., 2003, \mnras, 346, 1055 

\bibitem[\protect\citeauthoryear{Kehrig et al$.$}{2004}]{Kehrig2004} Kehrig, C., Telles, E., Cuisinier, F., 2004, \aj, 128, 1141 

\bibitem[\protect\citeauthoryear{Kewley et al$.$}{2001}]{Kewley2001} Kewley, L.J., Dopita, M.A., Sutherland, R.S., Heisler, C.A., Trevena, J., 2001, \apj, 556, 121

\bibitem[\protect\citeauthoryear{Kim et al$.$}{2017}]{Kim2017}  Kim, J., Chung, A., Wong, O.I., Lee, B., Sung, E., Staveley-Smith, L., 2017, \aap, 605, A54
 
\bibitem[\protect\citeauthoryear{Kunth et al$.$}{1985}]{Kunth1985} Kunth, D., Joubert, M., 1985, \aap, 142, 411

\bibitem[\protect\citeauthoryear{Kunth et al$.$}{1988}]{Kunth1988} Kunth, D., Maurogordato, S., Vigroux, L., 1988, \aap, 204, 10

\bibitem[\protect\citeauthoryear{Kunth et al$.$}{2000}]{Kunth2000}  Kunth, D., \"Ostlin, G., 2000, \aapr, 10, 1

\bibitem[\protect\citeauthoryear{Lagos et al$.$}{2007}]{Lagos2007}  Lagos, P., Telles, E., Melnick, J., 2007, \aap, 476, 89

\bibitem[\protect\citeauthoryear{Lagos et al$.$}{2009}]{Lagos2009} Lagos, P., Telles, E., Muñoz-Tuñón, C., Carrasco, E.R., Cuisinier, F., Tenorio-Tagle, G., 2009, \aj, 137, 5068 

\bibitem[\protect\citeauthoryear{Lagos et al$.$}{2011}]{Lagos2011} Lagos, P.,Telles, E., Nigoche-Netro, A., Carrasco, E.R., 2011, \aj, 142, 162

\bibitem[\protect\citeauthoryear{Lepine et al$.$}{2003}]{Lepine2003} Lepine, J.R.D., de Oliveira, A.C., Figueredo, M.V., Castilho, B.V., Gneiding, C., Barbuy, B. et al. 2003, SPIE, 4841, 1086

\bibitem[\protect\citeauthoryear{Luridiana et al$.$}{2015}]{Luridiana2015} Luridiana, V., Morisset, C., Shaw, R.A., 2015,  \aap, 573, A42

\bibitem[\protect\citeauthoryear{MacLow \& Ferrara}{1999}]{MacLow1999} Mac Low, M-M., Ferrara, A., 1999, \apj, 513, 142

\bibitem[\protect\citeauthoryear{McCall et al$.$}{1985}]{McCall1985} McCall, M.L., Rybski, P.M., Shields, G.A., 1985, \apjs, 57, 1

\bibitem[\protect\citeauthoryear{Masegosa et al$.$}{1994}]{Masegosa1994}  Masegosa, J., Moles, M., Campos-Aguilar, A., 1994, \apj, 420, 576
 
\bibitem[\protect\citeauthoryear{Marino et al$.$}{2013}]{Marino2013} Marino, R.A., Rosales-Ortega, F.F., Sánchez, S.F., Gil de Paz, A., Vílchez, J., Miralles-Caballero, D., Kehrig, C., Pérez-Montero et al. 2013, \aap, 559, 114

\bibitem[\protect\citeauthoryear{Markarian et al$.$}{1967}]{Markarian1967} Markarian, B.E.,  1967, Astrofizika, 3, 24

\bibitem[\protect\citeauthoryear{Méndez \& Esteban et al$.$}{2000}]{Mendez2000} Méndez, D.I.,  Esteban, C., 2000, \aap, 359, 493

\bibitem[\protect\citeauthoryear{Moiseev \& Lozinskaya}{2012}]{Moiseev2012} Moiseev, A.V., Lozinskaya, T.A., 2012 \mnras, 423, 1831 

\bibitem[\protect\citeauthoryear{Muñoz-Tuñón et al$.$}{1996}]{MunozTunon1996} Muñoz-Tuñón, C., Tenorio-Tagle, G., Castañeda, H.O., Terlevich, R., 1996, \aj, 112, 1636

\bibitem[\protect\citeauthoryear{Muñoz-Mateos et al$.$}{2009}]{MunozMateos2009} Muñoz-Mateos, J.C., Gil de Paz, A., Zamorano, J., Boissier, S., Dale, D. A., Pérez-González, P.G., Gallego, J., Madore et al. 2009, \apj, 703, 1569

\bibitem[\protect\citeauthoryear{Oparin \& Moiseev}{2018}]{Oparin2018}  Oparin, D.V., Moiseev, A.V., 2018, Astrophysical Bulletin, Vol. 73, Issue 3, p298

\bibitem[\protect\citeauthoryear{Osterbrock \& Ferland}{2006}]{Osterbrock2006} Osterbrock, D.E., Ferland, G.J., 2006, Astrophysics of gaseous nebulae and active galactic nuclei, 2nd. ed. by D.E. Osterbrock and G.J. Ferland. Sausalito, CA: University Science Books, 

\bibitem[\protect\citeauthoryear{Papaderos et al$.$}{1996a}]{Papaderos1996a} Papaderos, P., Loose, H.-H., Thuan, T.X., Fricke, K.J., 1996,  \aaps, 120, 207

\bibitem[\protect\citeauthoryear{Papaderos et al$.$}{1996b}]{Papaderos1996b} Papaderos, P., Loose, H.-H., Fricke, K.J., Thuan, T.X., 1996, \aap, 314, 59

\bibitem[\protect\citeauthoryear{Papaderos et al$.$}{2002}]{Papaderos2002}Papaderos, P., Izotov, Y.I., Thuan, T.X., Noeske, K.G., Fricke, K.J., Guseva, N.G., Green, R.F., 2002, \aap, 393, 421 

\bibitem[\protect\citeauthoryear{Paturel et al$.$}{2003}]{Paturel2003} Paturel, G., Theureau, G., Bottinelli, L., Gougenheim, L., Courdreau-Durand, N.,
Hallet, N., Petit, C. 2003, \aap, 412, 57

\bibitem[\protect\citeauthoryear{Penston et al$.$}{1977}]{Penston1977} Penston, M.V., Fosbury, R.A., Ward, M., Wilson, A., 1977, \mnras, 180, 19

\bibitem[\protect\citeauthoryear{Perret et al$.$}{2014}]{Perret2014} Perret, V., Renaud, F., Epinat, B., Amram, P., Bournaud, F., Contini, T., Teyssier, R., Lambert, J.C., \aap, 562, 1

\bibitem[\protect\citeauthoryear{Pustilnik \& Martin}{2007}]{Pustilnik2007} Pustilnik, S.A., Martin, J.-M. 2007, \aap, 464, 859

\bibitem[\protect\citeauthoryear{Pilyugin et al$.$}{2012}]{Pilyugin2012} Pilyugin, L.S., Grebel, E.K., Mattsson, L., 2012, \mnras, 424, 2316

\bibitem[\protect\citeauthoryear{Pilyugin \& Grebel}{2016}]{Pilyugin2016} Pilyugin, L.S., Grebel, E.K., 2016, \mnras,  457, 3678

\bibitem[\protect\citeauthoryear{Reif et al$.$}{1982}]{Reif1982} Reif, K., Mebold, U., Goss, W.M., van Woerden, H., Siegman, B., 1982, \aaps, 50, 451

\bibitem[\protect\citeauthoryear{Rybicki \& Lightman}{2004}]{Rybicki2004} Rybicki, G.B., Lightman, A.P., Radiative Processes in Astrophysics, 2004, Ed. WILEY-VCH Verlag GmbH \& Co. KGaA. 

\bibitem[\protect\citeauthoryear{Russeil et al$.$}{2016}]{Russeil2016} D. Russeil, J. Tigé, C. Adami, L.D. Anderson, N. Schneider, A. Zavagno, M.R. Samal, P. Amram, et al., 2016, \aap, 587, A135

\bibitem[\protect\citeauthoryear{Salzer et al$.$}{1989a}]{Salzer1989a} Salzer, J.J., MacAlpine, G,M., Boroson, T.A., 1989, \apjs,  70, 447

\bibitem[\protect\citeauthoryear{Salzer et al$.$}{1989b}]{Salzer1989b} Salzer, J.J., MacAlpine, G,M., Boroson, T.A., 1989, \apjs,  70, 479

\bibitem[\protect\citeauthoryear{Sargent et al$.$}{1970}]{Sargent1970}  Sargent, W.L.W., Searle, L., 1970, \apj, 162, L155

\bibitem[\protect\citeauthoryear{Schaerer et al$.$}{1999}]{Schaerer1999} Schaerer, D., Contini, T., Pindao, M., 1999,  \aaps, 136, 35

\bibitem[\protect\citeauthoryear{Schwartz et al$.$}{2006}]{Schwartz2006} Schwartz, C.M., Martin, C.L., Chandar, R., Leitherer, C., Heckman, T.M., Oey, M.S., 2006, \apj, 646, 858

\bibitem[\protect\citeauthoryear{Shaw et al$.$}{1995}]{Shaw1995}  Shaw, R.A., Dufour, R.J., 1995, \pasp, 107, 896

\bibitem[\protect\citeauthoryear{Smith}{1975}]{Smith1975} Smith, M.G, 1975, \apj, 202, 591

\bibitem[\protect\citeauthoryear{Smith et al$.$}{1976}]{Smith1976} Smith, M.G., Aguirre, C., Zemelman, M., 1976, \apjsupp, 32, 217S

\bibitem[\protect\citeauthoryear{Stevens et al$.$}{2002}]{Stevens2002} Stevens, I.R., Forbes, D.A., Norris, R.P., 2002, \mnras,  335, 1079

\bibitem[\protect\citeauthoryear{Sung et al$.$}{2002}]{Sung2002} Sung, E.C., Chun, M.S., Freeman, K. C., Chaboyer, B., in The Dynamics, Structure \& History of Galaxies: A Workshop in Honour of Professor Ken Freeman. ASP Conference Proceedings, Vol. 273, 2002, p.341

\bibitem[\protect\citeauthoryear{Telles \& Terlevich}{1995}]{Telles1995}  Telles, E., Terlevich, R., 1995, \mnras, 275, 1

\bibitem[\protect\citeauthoryear{Telles \& Terlevich}{1997a}]{Telles1997a} Telles, E., Terlevich, R., 1997, \mnras, 286, 183 

\bibitem[\protect\citeauthoryear{Telles et al$.$}{1997b}]{Telles1997b} Telles, E. Melnick, J., Terlevich, R., 1997, \mnras, 288, 78

\bibitem[\protect\citeauthoryear{Terlevich et al$.$}{1991}]{Terlevich1991} Terlevich, R., Melnick, J., Masegosa, J., Moles, M., Copetti, M.V.F., 1991, \aaps, 91, 285 

\bibitem[\protect\citeauthoryear{Terlevich \& Melnick}{1981}]{Terlevich1981}  Terlevich, R., Melnick, J., 1981, \mnras, 195, 839

\bibitem[\protect\citeauthoryear{Tody}{1986}]{Tody1986} Tody, D. 1986, SPIE, 627, 733

\bibitem[\protect\citeauthoryear{Thuan \& Martin}{1981}]{Thuan1981} Thuan, T.X., Martin, G.E., 1981, \apj, 247, 823

\bibitem[\protect\citeauthoryear{Thuan}{1983}]{Thuan1983} Thuan, T.X., 1983, \apj, 268, 667

\bibitem[\protect\citeauthoryear{Thuan}{1985}]{Thuan1985} Thuan, T.X. 1985, \apj, 299, 881

\bibitem[\protect\citeauthoryear{Torres-Campos et al$.$}{2017}]{Torres2017} Torres-Campos, A., Terlevich, E., Rosa-González, D., Terlevich, R., Telles, E., Díaz, A.I., 2017, \mnras, 471, 2829

\bibitem[\protect\citeauthoryear{Vacca \& Conti}{1992}]{Vacca1992} Vacca, W.D., Conti, P.S., 1992, \apj, 401, 543

\bibitem[\protect\citeauthoryear{Vanzi et al$.$}{2011}]{Vanzi2011} Vanzi, L., Cresci, G., Sauvage, M., Thompson, R., 2011, \aap, 534, A70

\bibitem[\protect\citeauthoryear{Veilleux \& Osterbrock}{1987}]{Veilleux1987} Veilleux, S., Osterbrock, D.E., 1987, \apjs, 63, 295

\bibitem[\protect\citeauthoryear{Westera et al$.$}{2004}]{Westera2004} Westera, P., Cuisinier, F., Telles, E., Kehrig, C., 2004, \mnras, 423, 133

\bibitem[\protect\citeauthoryear{Zhang et al$.$}{2020}]{Zhang2020} Zhang, Hong-Xin, Smith, R., Oh, Se-Heon, Paudel, S., Duc, P.A., Boselli, A., Côté, P., Ferrarese, L., et al. 2020, \apj, 900, 152

\bibitem[\protect\citeauthoryear{Zwicky}{1964}]{Zwicky1964} Zwicky, I.F., 1964, \apj, 140, 1467

\bibitem[\protect\citeauthoryear{Zwicky}{1966}]{Zwicky1966}  Zwicky, I.F., 1966, \apj, 143, 192

\bibitem[\protect\citeauthoryear{Zwicky \& Zwicky}{1971}]{Zwicky1971}  Zwicky, F., Zwicky, M.A., 1971, Guemligen: Zwicky, |c1971


\end{thebibliography}




\appendix

\section{Histograms of different parameters: [SII]$\lambda$6716/[SII]$\lambda$6731, EW(H$\beta$), E(B-V) and diagnostics diagrams line ratios}

\begin{figure}
	\includegraphics[width=0.5\textwidth]{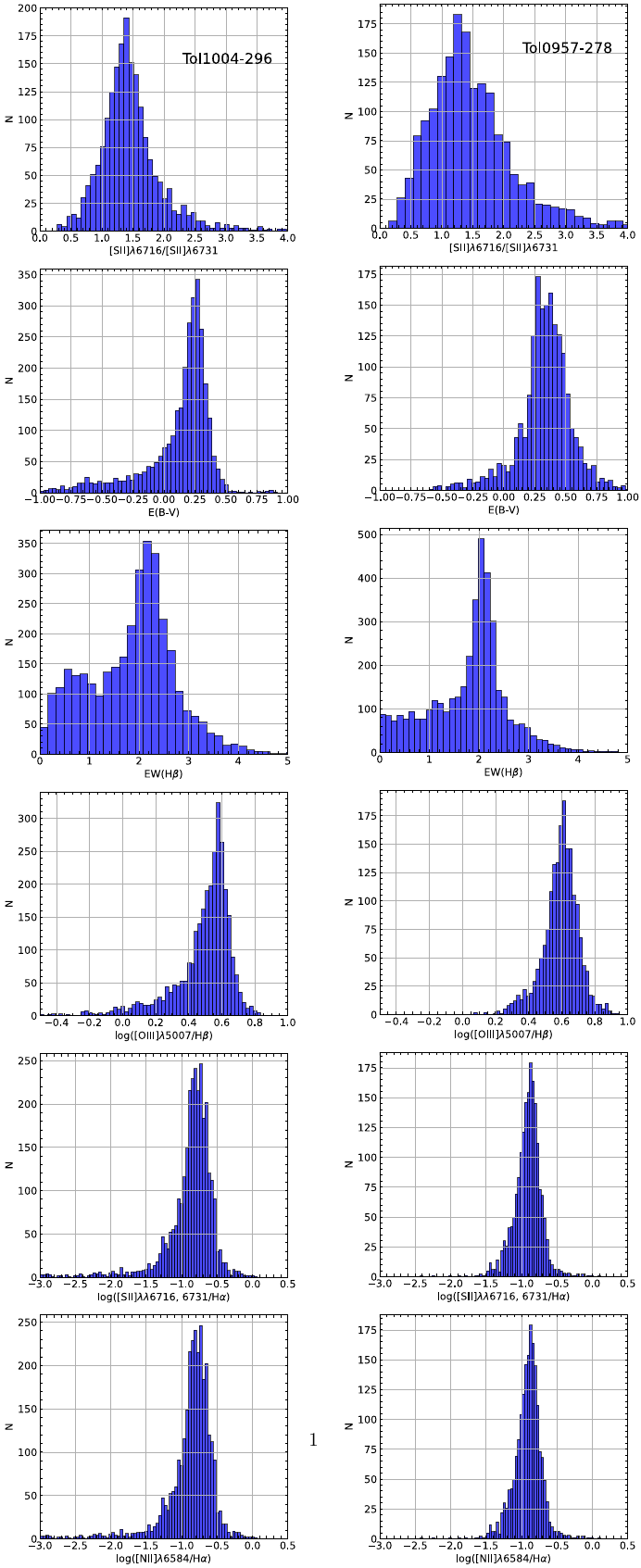}
    \caption{Histograms representing different parameters: [SII]$\lambda$6716/[SII]$\lambda$6731, EW(H$\beta$), E(B-V) and diagnostics diagrams line ratios for both galaxies, left column for Tol 1004-296 and right column for Tol 0957-278}.  
    \label{fig:Histograms}
\end{figure} 

\section{Tol 1004-296 MUSE data}

\subsection{Monochromatic Maps}

\begin{figure*}
	\includegraphics[width=\textwidth]{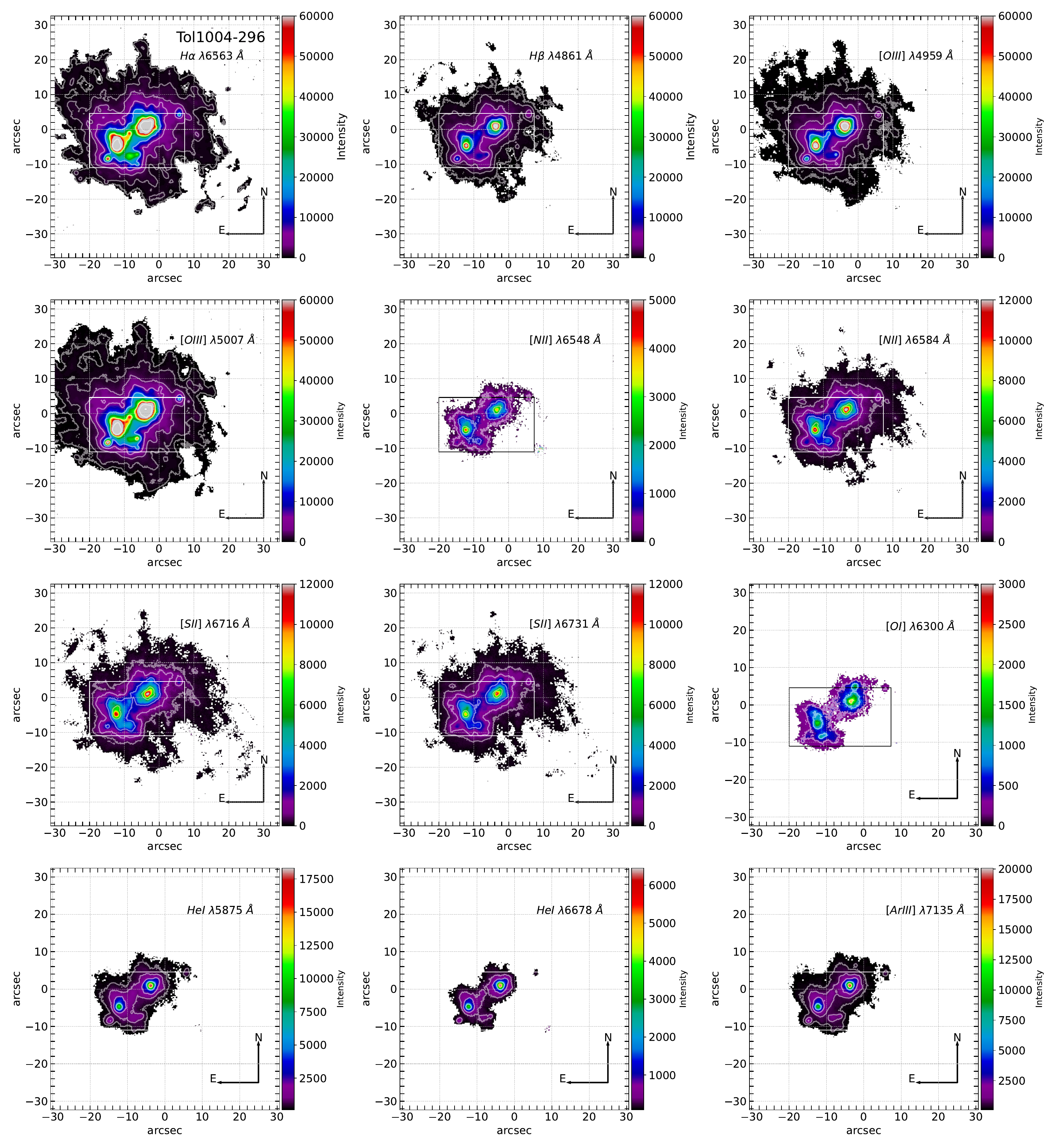}
    \caption{Tol 1004-296 MUSE data - Monochromatic Maps. Rectangle represents SIFS FoV.} 
    \label{fig:MonoMapsTol1004-296MUSE}
\end{figure*} 

\subsection{Lines Diagnostics maps}

\begin{figure}
	\includegraphics[width=0.4\textwidth ]{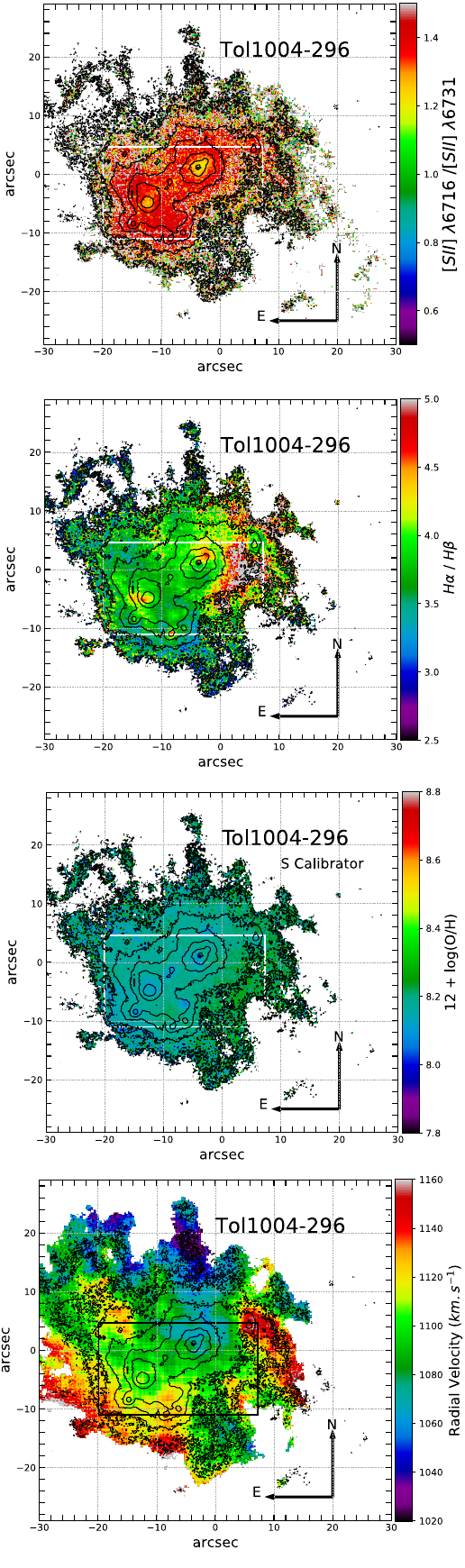}
    \caption{Tol 1004-296 MUSE data: From Top to Bottom: [SII]6716\AA / [SII]6731\AA map (electron Density); $H\alpha/H\beta$ map (Extinction); Metallicity map (S Calibrator); Radial Velocity Map. Rectangle represents SIFS FoV.} 
    \label{fig:LinesDisgnosticVelocityMapsTol1004-296MUSE}
\end{figure} 

\subsection{Ionization mechanisms diagrams and Velocity map}

\begin{figure}
	\includegraphics[width=0.328\textwidth]{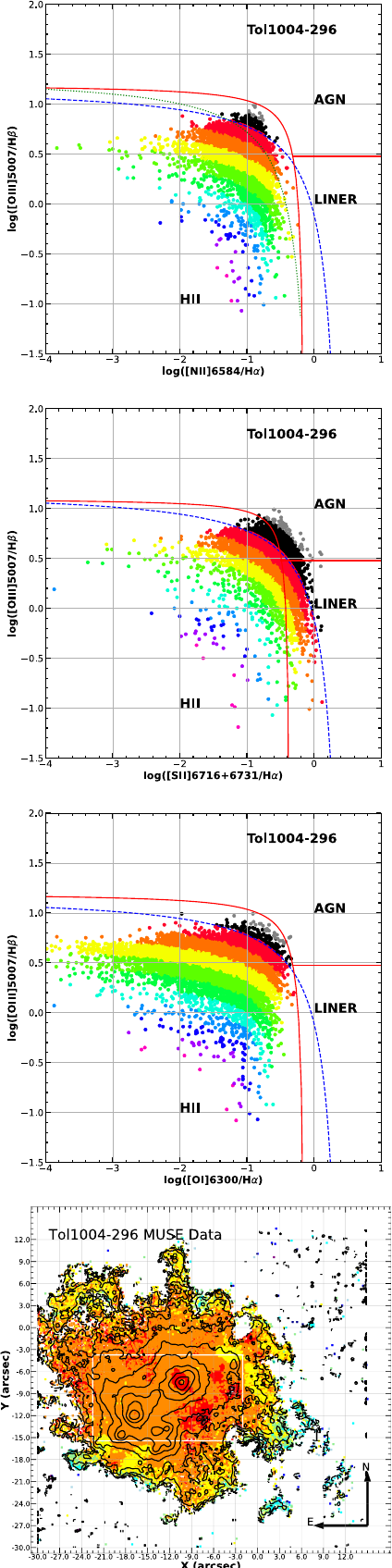}
    \caption{BPT diagrams for Tol 1004-296 using the MUSE data. Colors represent the distance from the \citet{Kewley2001} separation (blue dash line). Red lines represent separation from \citet{Veilleux1987}. 
    Bottom map represents the spatial location of the different points from the $[OIII]5007/H\beta vs [NII]6584/H\alpha$ diagram. Rectangle represents SIFS FoV.} 
    \label{fig:BPT_PlotTol1004-296MUSE}
\end{figure} 



\bsp	
\label{lastpage}
\end{document}